\documentclass{emulateapj}
\shorttitle{A Spitzer-IRS Survey of Mid-Infrared Molecular Emission II}
\shortauthors{Salyk et al.}
\newcounter{minirefcount}
\usepackage{amssymb}
\usepackage{amsmath}
\usepackage{multirow}
\usepackage{slashbox}
\usepackage{wasysym}
\usepackage{rotating}
\usepackage{natbib}
\newcommand{\miniref}[2]{\refstepcounter{minirefcount}\label{#2}(\arabic{minirefcount}) #1}
\newcommand{\water}{H$_2$O}
\newcommand{\acetylene}{C$_2$H$_2$}

\begin{document}
\title{A Spitzer Survey of Mid-Infrared Molecular Emission from Protoplanetary Disks II: Correlations and LTE Models \footnote[1]{Published in the Astrophysical Journal: ApJ, 731, 130, http://stacks.iop.org/0004-637X/731/130 }}
\author{C. Salyk} 
\affil{McDonald Observatory, The University of Texas at Austin, 1 University Station, C1402, Austin, TX 78712, USA \footnote[2]{csalyk@astro.as.utexas.edu}}
\author{K.~M. Pontoppidan}
\affil{Space Telescope Science Institute, 3700 San Martin Drive, Baltimore, MD 21218, USA}
\author{G.~A. Blake}
\affil{Division of Geological \& Planetary Sciences, Mail Stop 150-21, California Institute of Technology, Pasadena, CA 91125, USA}
\author{J.~R. Najita}
\affil{National Optical Astronomy Observatory, 950 N. Cherry Ave., Tucson, AZ 85719, USA}
\author{J.~S. Carr}
\affil{Naval Research Laboratory, Code 7211, Washington, DC 20375, USA}

\begin{abstract}
We present an analysis of Spitzer-IRS observations of \water, OH, HCN, \acetylene\, and CO$_2$ emission, and Keck-NIRSPEC observations of CO emission, from a diverse sample of T Tauri and Herbig Ae/Be circumstellar disks.  We find that detections and strengths of most mid-IR molecular emission features are correlated with each other, suggesting a common origin and similar excitation conditions for this mid-infrared line forest.  Aside from the remarkable differences in molecular line strengths between T Tauri, Herbig Ae/Be and transitional disks discussed in \citet{Pontoppidan10}, we note that the line detection efficiency is anti-correlated with the 13/30 $\mu$m SED spectral slope, which is a measure of the degree of grain settling in the disk atmosphere.  We also note a correlation between detection efficiency and H$\alpha$ equivalent width, and tentatively with accretion rate, suggesting that accretional heating contributes to line excitation.
If detected, \water\ line fluxes are correlated with the mid-IR continuum flux, and other co-varying system parameters, such as $L_\star$.  However,  significant sample variation, especially in molecular line ratios, remains, and its origin has yet to be explained.  LTE models of the \water\ emission show that line strength is primarily related to the best-fit emitting area, and this accounts for most source-to-source variation in \water\ emitted flux.  Best-fit temperatures and column densities cover only a small range of parameter space, near $\sim10^{18}\,\mathrm{cm}^{-2}$ and 450 K for all sources, suggesting a high abundance of \water\ in many planet-forming regions.  Other molecules have a range of excitation temperatures from $\sim500-1500\,$K, also consistent with an origin in planet-forming regions.
We find molecular ratios relative to water of $\sim10^{-3}$ for all molecules, with the exception of CO, for which n(CO)/n(\water)$\sim$1.  However, LTE fitting caveats and differences in the way thermo-chemical modeling results are reported make comparisons with such models difficult, and highlight the need for additional observations coupled with the use of line-generating radiative transfer codes.

\end{abstract}
 
 \maketitle
\section{Introduction}
The chemistry of protoplanetary disks plays a pivotal role in the development of planetary systems, of habitable planets,
and of life.  However, our understanding of the chemical conditions in protoplanetary disks---especially in 
planet forming regions---is far from complete.  In particular, there are many 
uncertainties about the role of volatiles such as water, which not only play a crucial role in life as we know it, but also in
many other processes that affect the formation and evolution of planets.  For example, the location of water-ice sublimation, 
the so-called ice line, may define a boundary within which planetary cores do not grow large enough to become gas giants.
And, molecular ices make up the bulk of many outer solar system bodies, including Kuiper Belt objects
and outer-planet satellites. 

For many years, the detection of water in protoplanetary disks remained elusive, with only a few definitive examples of ice \citep{Malfait98, Terada07}
and vapor \citep{Najita00, Carr04, Thi05} detections.   Additionally, organic molecules were detected in absorption through an edge-on disk \citep{Lahuis06}.  
Recently, however, \citet{Carr08} and \citet{Salyk08} reported 
detections of a forest of emission lines of \water, OH, HCN, \acetylene\ and CO$_2$ in high-resolution spectra from the Spitzer Space Telescope
InfraRed Spectrograph (IRS), as well as of vibrational \water\ lines with the Keck Near Infrared Spectrograph (NIRSPEC).  Based on emitting temperatures, fluxes and line shapes, these lines appeared to originate in the disk atmospheres at terrestrial-planet-forming radii.  Additionally, when present, the mid-IR water lines likely dominate the cooling of the inner disk surface layer \citep{Pontoppidan10}.

The frequent detections of water and other molecules in the inner regions of protoplanetary disks, made possible largely by optimized high dynamic range observing strategies and data reduction routines \citep{Carr08}, mark a turning point in our ability to study chemical evolution and molecular transport in the inner  solar system.  
Because chemistry and line excitation are linked to disk physics, molecular emission can be used
to study disk structure and physics.  For example, water is destroyed by photodissociation and by high levels of ionization \citep[e.g.][]{Bergin03, Glassgold09}, but is capable of self-shielding in certain conditions \citep{Bethell09}, and so water abundances may reflect the disk irradiation environment.  Because water condenses
at relatively small radii in the disk midplane ($\sim$ a few AU for solar-mass stars), water vapor may be driven outward by vapor pressure differences,
while water ice is carried inwards by planetesimals.  Hence, water abundances may also reflect radial transport and planet formation timescales.  
Water abundances may similarly depend on vertical transport, which, when coupled with midplane condensation, could deplete water 
even well inside the canonical snow line \citep{Meijerink09}. Finally, the strength of molecular emission lines depends upon the disk structure as a whole, including the gas temperature structure and the degree of dust settling/grain growth. 

One way to disentangle the many factors that affect molecular abundances and line strengths is to observe a large sample of disks, and look for trends predicted by the processes described above.  Here we report characteristics of a sample of disks observed with the Spitzer Space Telescope Infrared Spectrograph (IRS), first described and presented in \citet{Pontoppidan10} (hereafter Paper I).  In Paper I, we presented the overall characteristics of the spectra, including molecular detection rates, and reported a strong dependence of molecular emission on spectral type.  In this paper, we analyze the IRS spectra in more detail, reporting correlations of line strengths and detections with system parameters, as well as molecular ratios calculated from LTE slab models.  In addition, we use observations of CO rovibrational emission at 5 $\mu$m (G.A. Blake et al., in prep), to estimate molecular ratios with respect to CO.   As one of the most abundant and easily-detected molecules in disks, CO is a tracer of the bulk of the disk gas, and provides a good baseline for comparing molecular abundances.  M-band ($\sim$5\,$\mu$m) CO rovibrational lines have similar excitation energies and collisional rates as the rotational \water\ lines observed with the IRS, making these
transitions particularly suited for comparison with the IRS observations.   

\section{Observations and Reduction}
\subsection{Spitzer-IRS}
The mid-IR data in this study were all obtained with the Spitzer Infrared Spectrograph (IRS) in its high-resolution mode.  The observed sample derives from a high S/N program designed for the detection of water and other molecular emission (PID 50641; PI J. Carr), as well as from a few programs in the Spitzer archive, and is described in detail in Paper I.   A description of the observations, including program IDs, integration times and Astronomical Observation Request (AOR) numbers, can also be found in Paper I.

A brief explanation of the data reduction procedure is as follows.  All data were reduced with our own IDL routines, along with the IRSFRINGE package \citep{Lahuis07},  using a procedure similar to that described in \citet{Carr08}.  We began with IRS droop-corrected, non-flat-fielded basic calibrated data (BCD) from the Spitzer Science Center IRS pipeline.  In order to reduce noise caused by edge-effects, pipeline flat-fields were first divided by a low-order polynomial fit in the spectral and spatial directions.  An off-frame was subtracted, data were flat-fielded, and noisy and bad pixels removed.  Spectra were divided by an average of at least five standard stars, and minimally processed with the IRSFRINGE package.  A more complete description of the reduction procedure can be found in Paper I.  

As reported in Paper I, the observed IRS spectra contain a forest of molecular emission lines from \water, OH, HCN, C$_2$H$_2$ and CO$_2$.   For the analyses in Paper I and this work, positive detections require a 3.5$\sigma$ peak (at two line locations for \water\ and OH), with some line detections included or excluded through individual by-eye examination.  Fluxes are calculated by defining and subtracting a linear fit to the local continuum, then summing over the expected width of the line, except for the water emission lines, which are fit with Gaussians, as described in Paper I.   Wavelength regions used are as follows: OH (23.009--23.308 and 27.308--27.764 $\mu$m), HCN (13.837--14.075 $\mu$m), C$_2$H$_2$ (13.553--13.764 $\mu$m) and CO$_2$ (14.847--15.014 $\mu$m).   All features overlap to some extent with \water\ emission, with flux contributions from water as high as $\sim40$\% in the HCN and C$_2$H$_2$ regions, and even higher near the CO$_2$ Q-branch.  Thus, an \water\ emission model (described in \S \ref{sec:lte}) was subtracted before molecular line fluxes were computed.  Nevertheless, this subtraction does not have a significant effect on the statistical analyses we present.  

Statistical errors on individual points in the spectrum are calculated as $\sigma/\sqrt{N}$, where $N$ is the number of frames used to calculate the flux, and $\sigma$ is the standard deviation of the values in that pixel.  If $N<3$, then the error is taken to be the standard deviation computed using three neighboring pixels in the dispersion direction.  Systematic uncertainties are undoubtedly larger, especially for frames without dedicated background observations, but can be difficult to characterize. 

Line flux errors are taken to be the root squared sum of errors on individual points over the wavelength range used to compute the line flux.  Upper limits are calculated assuming a 3.5$\sigma$ line height (see Paper I for a discussion of this choice), and a feature shape defined by an LTE slab model, with column density, $N$, and temperature, $T$, taken to be near the mean of those derived in \S \ref{sec:lte}.  However, upper limits are not very sensitive to the model parameters.  Calculated fluxes and upper limits for all molecules except for water are shown in Tables \ref{table:fluxes} and \ref{table:fluxes_haebe}.  Detections and water fluxes can be found in Paper I.

We also compute the flux in the 12.5--14 $\mu$m interval (F$_{13}$) and the 13/30 $\mu$m SED slope ($n_{13-30}$), which are shown in Tables \ref{table:star_refs} and \ref{table:herbig}.  Precise definitions of these parameters can be found in \citet{Furlan06}. Statistical errors are small; however, note that the variation in flux over the 12.5--14 $\mu$m interval can be large (as much as several Jy), and that systematic errors can arise from background subtraction and the computation of the spectral response function.

\begin{deluxetable*}{lccccccc}
\tabletypesize{\scriptsize}
\tablecaption{Line Fluxes: T Tauri Disks  \label{table:fluxes}}
\startdata
 \hline
                Name &         OH &         OH &        HCN & C$_2$H$_2$ &     CO$_2$ &         CO &         CO  \\
                     &  (23.2 $\mu$m) &  (27.6 $\mu$m) &                &                &                &           P(6) &          P(12)  \\
\hline \\
            LkHa 270 &           $<$ 1.35 &           $<$ 1.35 &           $<$ 2.15 &           $<$ 1.37 &    2.37 $\pm$ 0.24 &                 ns &                 ns  \\
            LkHa 271 &           $<$ 1.26 &           $<$ 1.28 &           $<$ 1.06 &           $<$ 0.72 &           $<$ 0.56 &                 ns &                 ns  \\
            LkHa 326 &    1.56 $\pm$ 0.11 &    3.78 $\pm$ 0.12 &    0.65 $\pm$ 0.12 &           $<$ 0.55 &    1.49 $\pm$ 0.10 &    1.21$\pm$   0.1 &    1.30$\pm$   0.1  \\
            LkHa 327 &    4.15 $\pm$ 0.41 &    6.66 $\pm$ 0.44 &           $<$ 5.28 &           $<$ 3.52 &    5.33 $\pm$ 0.59 &    2.61$\pm$   0.1 &    2.60$\pm$   0.3  \\
            LkHa 330 &           $<$ 2.52 &           $<$ 2.76 &           $<$ 1.15 &           $<$ 0.76 &           $<$ 0.68 &    1.19$\pm$   0.0 &    1.05$\pm$   0.0  \\
              LkCa 8 &           $<$ 1.21 &           $<$ 1.21 &           $<$ 1.15 &           $<$ 0.71 &           $<$ 0.55 &                 ns &    0.65$\pm$   0.1  \\
              IQ Tau &           $<$ 1.43 &           $<$ 1.39 &    5.70 $\pm$ 0.30 &    4.12 $\pm$ 0.23 &           $<$ 1.02 &                 ea &                 ea  \\
            V710 Tau &           $<$ 0.92 &           $<$ 0.93 &    2.59 $\pm$ 0.17 &    1.55 $\pm$ 0.13 &    0.94 $\pm$ 0.13 &    0.87$\pm$   0.1 &    0.68$\pm$   0.1  \\
              AA Tau &    3.11 $\pm$ 0.10 &    2.76 $\pm$ 0.11 &    6.50 $\pm$ 0.16 &    2.92 $\pm$ 0.12 &    1.71 $\pm$ 0.12 &    0.88$\pm$   0.1 &    0.87$\pm$   0.1  \\
          CoKu Tau/4 &           $<$ 1.09 &           $<$ 1.09 &           $<$ 1.74 &           $<$ 1.13 &           $<$ 0.92 &         $<$   0.40 &         $<$   0.10  \\
              DN Tau &           $<$ 1.34 &           $<$ 1.33 &           $<$ 1.48 &           $<$ 0.99 &           $<$ 0.76 &         $<$   0.39 &         $<$   0.11  \\
              FX Tau &           $<$ 1.30 &           $<$ 1.27 &    0.88 $\pm$ 0.28 &           $<$ 1.27 &           $<$ 0.97 &                  a &                  a  \\
              DR Tau &    3.74 $\pm$ 0.30 &   14.79 $\pm$ 0.34 &   13.09 $\pm$ 0.53 &    6.00 $\pm$ 0.41 &    2.48 $\pm$ 0.43 &   10.40$\pm$   0.1 &   11.54$\pm$   0.6  \\
              SX Cha &    1.67 $\pm$ 0.09 &    3.92 $\pm$ 0.11 &    0.59 $\pm$ 0.16 &           $<$ 0.72 &           $<$ 0.58 &            \nodata &            \nodata  \\
              SY Cha &           $<$ 0.35 &           $<$ 0.36 &    1.22 $\pm$ 0.07 &    0.88 $\pm$ 0.05 &    0.51 $\pm$ 0.05 &            \nodata &            \nodata  \\
              TW Cha &    1.64 $\pm$ 0.04 &    1.99 $\pm$ 0.05 &    2.27 $\pm$ 0.06 &    1.36 $\pm$ 0.05 &           $<$ 0.22 &            \nodata &            \nodata  \\
              VW Cha &    8.22 $\pm$ 0.12 &   12.61 $\pm$ 0.14 &    4.37 $\pm$ 0.20 &    2.68 $\pm$ 0.16 &    2.48 $\pm$ 0.17 &            \nodata &            \nodata  \\
              VZ Cha &    2.56 $\pm$ 0.06 &    2.79 $\pm$ 0.07 &    3.69 $\pm$ 0.14 &    4.59 $\pm$ 0.10 &           $<$ 0.48 &            \nodata &            \nodata  \\
              WX Cha &    2.01 $\pm$ 0.06 &    2.05 $\pm$ 0.07 &    4.18 $\pm$ 0.11 &    2.68 $\pm$ 0.08 &    1.24 $\pm$ 0.09 &            \nodata &            \nodata  \\
              XX Cha &    0.64 $\pm$ 0.06 &    1.99 $\pm$ 0.07 &    1.20 $\pm$ 0.09 &    1.85 $\pm$ 0.06 &    0.86 $\pm$ 0.07 &            \nodata &            \nodata  \\
               T Cha &           $<$ 2.33 &           $<$ 2.52 &           $<$ 1.90 &           $<$ 1.24 &           $<$ 0.92 &            \nodata &            \nodata  \\
               Sz 50 &           $<$ 0.51 &           $<$ 0.50 &           $<$ 0.63 &           $<$ 0.41 &    1.23 $\pm$ 0.08 &            \nodata &            \nodata  \\
          HD 135344 B &           $<$ 2.49 &           $<$ 2.76 &           $<$ 1.93 &           $<$ 1.22 &           $<$ 0.95 &                 ns &    2.29$\pm$   0.1  \\
              HT Lup &           $<$ 3.29 &           $<$ 3.11 &           $<$ 9.25 &           $<$ 5.60 &    5.58 $\pm$ 1.08 &            \nodata &            \nodata  \\
              GW Lup &           $<$ 0.94 &           $<$ 0.96 &           $<$ 1.20 &           $<$ 0.74 &    2.08 $\pm$ 0.13 &         $<$   1.57 &         $<$   0.28  \\
              GQ Lup &    3.26 $\pm$ 0.12 &    5.52 $\pm$ 0.14 &    3.90 $\pm$ 0.16 &           $<$ 0.72 &           $<$ 0.56 &    5.32$\pm$   0.3 &    6.79$\pm$   0.6  \\
              IM Lup &           $<$ 0.87 &           $<$ 0.89 &           $<$ 1.22 &           $<$ 0.81 &    0.86 $\pm$ 0.15 &                 ns &                 ns  \\
           HD 142527 &           $<$14.58 &           $<$15.78 &           $<$11.17 &           $<$ 7.53 &           $<$ 5.77 &            \nodata &            \nodata  \\
              RU Lup &           $<$ 1.89 &           $<$ 1.76 &    3.69 $\pm$ 0.39 &           $<$ 1.81 &    1.41 $\pm$ 0.31 &            \nodata &            \nodata  \\
              RY Lup &           $<$ 2.81 &           $<$ 2.73 &           $<$ 3.55 &           $<$ 2.30 &           $<$ 1.83 &            \nodata &            \nodata  \\
              EX Lup &   14.43 $\pm$ 0.17 &    7.41 $\pm$ 0.19 &           $<$ 2.15 &           $<$ 1.48 &           $<$ 1.12 &            \nodata &            \nodata  \\
              AS 205 &   16.34 $\pm$ 0.58 &   47.74 $\pm$ 0.68 &   22.02 $\pm$ 1.95 &   21.09 $\pm$ 1.50 &   23.26 $\pm$ 1.60 &   24.18$\pm$   0.6 &   24.03$\pm$   1.6  \\
            Haro 1-1 &           $<$ 1.85 &           $<$ 1.87 &           $<$ 1.37 &           $<$ 0.89 &           $<$ 0.69 &         $<$   0.68 &            \nodata  \\
            Haro 1-4 &           $<$ 2.17 &           $<$ 2.13 &    2.18 $\pm$ 0.32 &    2.97 $\pm$ 0.24 &           $<$ 1.11 &                 ns &            \nodata  \\
               VSSG1 &           $<$ 3.51 &           $<$ 3.76 &   13.72 $\pm$ 0.54 &   17.42 $\pm$ 0.41 &    4.91 $\pm$ 0.40 &                 ea &                 ea  \\
            DoAr 24E &    5.72 $\pm$ 0.34 &   20.61 $\pm$ 0.40 &   30.60 $\pm$ 0.35 &   32.55 $\pm$ 0.27 &    3.47 $\pm$ 0.27 &   13.58$\pm$   0.5 &   14.53$\pm$   0.9  \\
             DoAr 25 &           $<$ 1.42 &           $<$ 1.48 &    4.88 $\pm$ 0.17 &    0.99 $\pm$ 0.13 &           $<$ 0.59 &                  a &                  a  \\
               SR 21 &           $<$ 8.38 &           $<$ 8.74 &           $<$ 5.32 &           $<$ 3.39 &           $<$ 3.01 &    0.47$\pm$   0.0 &    0.54$\pm$   0.0  \\
                SR 9 &           $<$ 1.62 &           $<$ 1.58 &           $<$ 3.14 &           $<$ 2.01 &           $<$ 1.61 &    0.77$\pm$   0.1 &    1.05$\pm$   0.2  \\
            V853 Oph &    1.99 $\pm$ 0.13 &    0.32 $\pm$ 0.14 &    1.85 $\pm$ 0.24 &    0.93 $\pm$ 0.18 &           $<$ 0.82 &    1.37$\pm$   0.1 &    1.49$\pm$   0.1  \\
             ROX 42C &           $<$ 2.05 &           $<$ 1.97 &           $<$ 2.17 &           $<$ 1.39 &           $<$ 1.11 &         $<$   1.91 &            \nodata  \\
            ROX 43 A &           $<$ 2.45 &           $<$ 2.22 &           $<$ 4.73 &           $<$ 3.10 &           $<$ 2.44 &         $<$   2.32 &            \nodata  \\
           Haro 1-16 &    5.80 $\pm$ 0.16 &    7.46 $\pm$ 0.17 &    3.23 $\pm$ 0.17 &           $<$ 0.82 &           $<$ 0.65 &    1.52$\pm$   0.2 &    2.40$\pm$   0.2  \\
           Haro 1-17 &           $<$ 1.00 &           $<$ 1.03 &           $<$ 0.99 &           $<$ 0.64 &           $<$ 0.50 &         $<$   1.00 &            \nodata  \\
              RNO 90 &    1.20 $\pm$ 0.38 &   25.60 $\pm$ 0.42 &   40.45 $\pm$ 0.65 &   17.77 $\pm$ 0.50 &    9.01 $\pm$ 0.54 &   20.95$\pm$   0.7 &   22.07$\pm$   0.7  \\
            Wa Oph 6 &    4.61 $\pm$ 0.13 &    4.23 $\pm$ 0.15 &    1.71 $\pm$ 0.20 &           $<$ 0.89 &    2.56 $\pm$ 0.17 &    3.66$\pm$   0.5 &    3.37$\pm$   0.5  \\
           V1121 Oph &           $<$ 3.89 &           $<$ 3.52 &           $<$ 5.76 &           $<$ 3.79 &           $<$ 3.02 &    4.38$\pm$   0.5 &    2.40$\pm$   0.4  \\
               EC 82 &           $<$ 1.48 &           $<$ 1.44 &           $<$ 1.56 &           $<$ 1.06 &           $<$ 0.88 &    4.47$\pm$   0.1 &    4.69$\pm$   0.1  \\
\enddata
\tablecomments{Fluxes are $10^{-14} \mathrm{erg\ cm}^{-2}\ \mathrm{s}^{-1}$.  Upper limits are 3.5$\sigma$; error bars are 1 $\sigma$.  For CO lines, {\it ns} means the line
is in emission but S/N or line coverage was not sufficient for determining a line flux.  {\it ea} means the line is in emission, but a significant absorption component makes it difficult to determine the line flux.  {\it a} means the line is in absorption.  An ellipsis means the source was not observed at this wavelength. }
\end{deluxetable*}

\begin{deluxetable*}{lccccccc}
\tabletypesize{\scriptsize}
\tablecaption{Line Fluxes: Herbig Ae/Be Disks  \label{table:fluxes_haebe}}
\startdata
 Herbig Ae/Be disks \\
 \hline
                Name &         OH &         OH &        HCN & C$_2$H$_2$ &     CO$_2$ &         CO &         CO  \\
                     &  (23.2 $\mu$m) &  (27.6 $\mu$m) &                &                &                &           P(6) &          P(12)  \\
                     \hline
            HD 36112 &           $<$11.19 &           $<$10.93 &           $<$ 8.43 &           $<$ 5.61 &           $<$ 4.52 &    3.80$\pm$   0.2 &    5.13$\pm$   0.1  \\
           HD 244604 &           $<$ 2.99 &           $<$ 2.73 &           $<$ 4.68 &           $<$ 3.20 &           $<$ 2.44 &    0.66$\pm$   0.1 &    0.43$\pm$   0.1  \\
            HD 36917 &           $<$ 3.71 &           $<$ 4.11 &           $<$ 5.72 &           $<$ 3.83 &           $<$ 2.82 &                 ns &    3.57$\pm$   0.6  \\
            HD 37258 &           $<$ 2.36 &           $<$ 2.10 &           $<$ 4.97 &           $<$ 3.07 &           $<$ 2.50 &         $<$   1.14 &         $<$   0.23  \\
              BF Ori &           $<$ 2.28 &           $<$ 2.10 &           $<$ 3.50 &           $<$ 2.31 &           $<$ 1.75 &         $<$   0.77 &         $<$   0.20  \\
            HD 37357 &           $<$ 3.41 &           $<$ 3.24 &           $<$ 4.23 &           $<$ 2.82 &           $<$ 2.20 &    0.33$\pm$   0.0 &    0.67$\pm$   0.1  \\
            HD 37411 &           $<$ 3.03 &           $<$ 3.00 &           $<$ 2.02 &           $<$ 1.32 &           $<$ 1.05 &         $<$   0.79 &         $<$   0.28  \\
              RR Tau &           $<$ 2.75 &           $<$ 2.67 &           $<$ 3.86 &           $<$ 2.58 &           $<$ 1.95 &                 ns &    0.61$\pm$   0.2  \\
            HD 37806 &           $<$ 8.64 &           $<$ 7.53 &           $<$11.70 &           $<$ 7.98 &           $<$ 5.98 &    1.68$\pm$   0.5 &    2.80$\pm$   0.3  \\
            HD 38087 &           $<$ 2.26 &           $<$ 2.53 &           $<$ 1.54 &           $<$ 1.08 &           $<$ 0.85 &            \nodata &            \nodata  \\
            HD 38120 &           $<$13.87 &           $<$12.79 &           $<$11.37 &           $<$ 7.64 &           $<$ 6.22 &    0.49$\pm$   0.0 &    0.28$\pm$   0.0  \\
            HD 50138 &           $<$22.98 &           $<$19.22 &           $<$27.07 &           $<$18.94 &           $<$13.87 &    6.72$\pm$   1.1 &   13.86$\pm$   1.2  \\
            HD 72106 &           $<$ 3.69 &           $<$ 3.48 &           $<$ 4.63 &           $<$ 3.14 &           $<$ 2.44 &         $<$   0.85 &         $<$   0.13  \\
            HD 95881 &           $<$ 6.18 &           $<$ 5.52 &           $<$10.25 &           $<$ 7.05 &           $<$ 5.22 &            \nodata &            \nodata  \\
            HD 98922 &           $<$13.65 &           $<$11.83 &           $<$27.54 &           $<$19.18 &           $<$14.06 &            \nodata &            \nodata  \\
           HD 101412 &           $<$ 2.57 &           $<$ 2.39 &           $<$ 4.92 &           $<$ 3.26 &   10.33 $\pm$ 0.55 &            \nodata &            \nodata  \\
           HD 144668 &           $<$12.01 &           $<$10.85 &           $<$14.19 &           $<$ 9.67 &           $<$ 7.24 &            \nodata &            \nodata  \\
           HD 149914 &           $<$ 1.00 &           $<$ 1.05 &           $<$ 1.19 &           $<$ 0.81 &           $<$ 0.55 &         $<$   0.81 &         $<$   0.17  \\
           HD 150193 &           $<$13.84 &           $<$12.21 &           $<$13.20 &           $<$ 8.92 &           $<$ 6.93 &    6.98$\pm$   0.4 &    4.74$\pm$   0.3  \\
              VV Ser &           $<$ 3.24 &           $<$ 2.90 &           $<$ 5.37 &           $<$ 3.69 &           $<$ 2.77 &    3.04$\pm$   0.3 &    2.68$\pm$   0.3  \\
            LkHa 348 &           $<$ 5.26 &           $<$ 4.49 &           $<$11.24 &           $<$ 7.70 &           $<$ 5.60 &                  a &                  a  \\
           HD 163296 &           $<$ 9.01 &           $<$ 8.08 &           $<$13.74 &           $<$ 9.38 &           $<$ 7.22 &    8.50$\pm$   1.2 &    9.06$\pm$   3.1  \\
           HD 179218 &           $<$19.55 &           $<$18.59 &           $<$16.96 &           $<$11.68 &           $<$ 9.14 &                 ns &    2.33$\pm$   0.4  \\
           HD 190073 &           $<$ 5.76 &           $<$ 5.14 &           $<$ 9.87 &           $<$ 6.56 &           $<$ 4.95 &    4.15$\pm$   1.3 &    3.00$\pm$   0.7  \\
            LkHa 224 &           $<$ 5.37 &           $<$ 5.68 &           $<$ 8.66 &           $<$ 5.88 &           $<$ 4.50 &                 ns &                 ns  \\
\enddata
\tablecomments{See Notes for Table \ref{table:fluxes}}
\end{deluxetable*}

\subsection{Keck-NIRSPEC}
M-band ($\sim$5 $\mu$m) spectra of CO $\Delta v$=1 rovibrational emission were obtained for 54 of the 73 sources presented here.  The data were obtained as part of a large M-band survey of Herbig Ae/Be (HAeBe) and classical T Tauri star (cTTs) disks, obtained from 2000--2010 with NIRSPEC \citep{McLean00} on the Keck II telescope.  The spectra will be described in detail in G.A. Blake et al. (in preparation).  However, we provide a few key results in this work, so as to compare with the Spitzer-IRS emission lines. Detailed analysis has been previously presented for transitional disks \citep{Salyk08,Salyk09} and a few HAeBe disks \citep{Blake04}, representing only a subset of the full survey.  Line fluxes for the P(6) and P(12) transitions are shown in Tables \ref{table:fluxes} and \ref{table:fluxes_haebe}.  Upper limits are 3$\sigma$ and assume a line width of 5$\times10^{-4}$ $\mu$m ($\sim30$ km s$^{-1}$).  

\section{Correlations}
\subsection{Detection rate correlations between molecules}
\begin{figure}
\epsscale{1}
\plotone{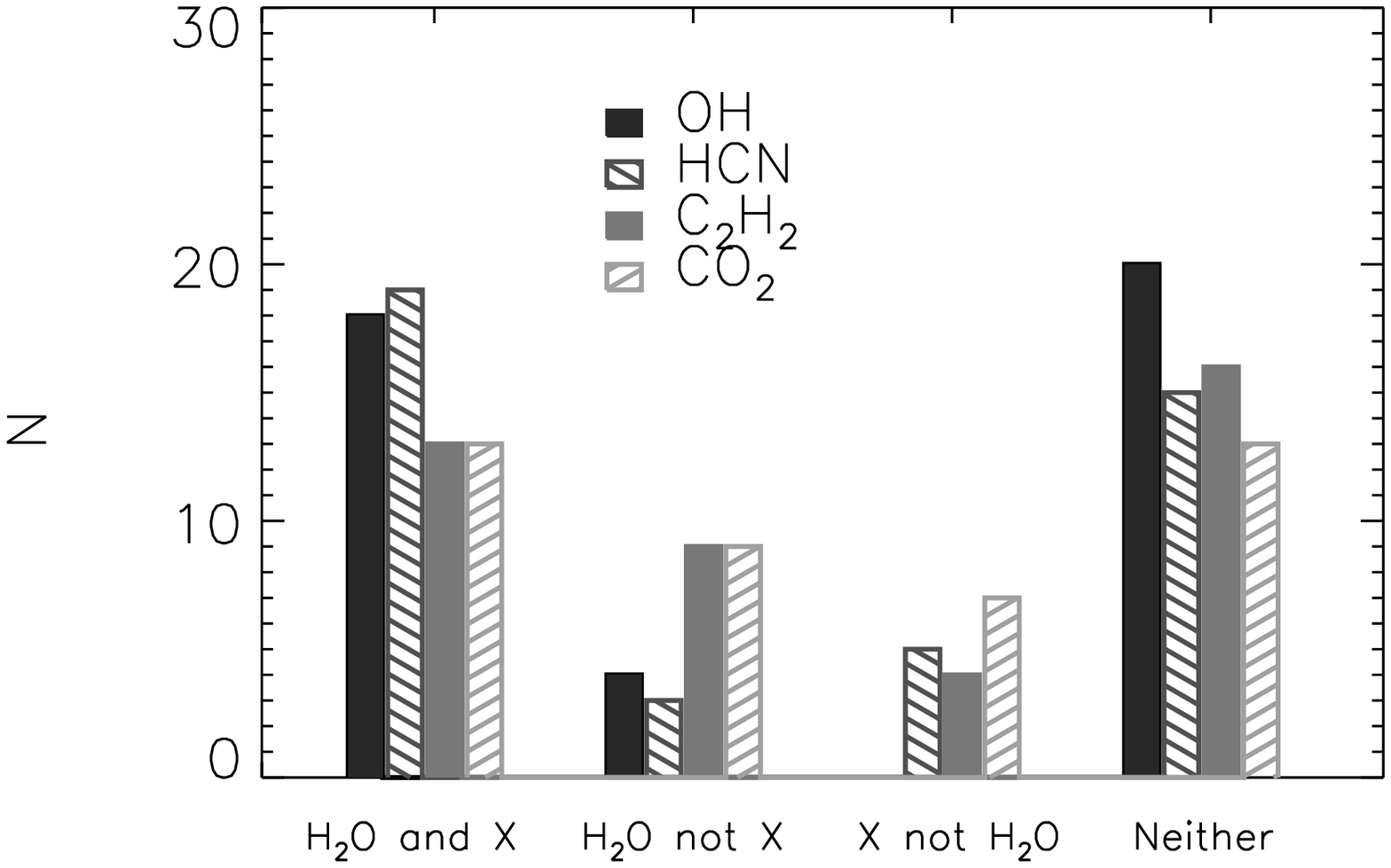}
\caption{Number of cTTs for which \water\ and molecule X are both detected, or only one is detected, or neither is detected.  (HAeBe and transitional disks are excluded for clarity, since all are non-detections).
 \label{fig:detect_hist}}
\end{figure}

Detections of most molecules are strongly correlated with each other, even when HAeBe and transitional disk non-detections (discussed in Paper I) are excluded from consideration.  In Figure \ref{fig:detect_hist} we show the number of cTTs in each of four categories: both \water\ and molecule X are detected, \water\ is detected but molecule X  is not, molecule X is detected but \water\ is not, and neither molecule is detected.  For all molecules it is much more common for both or neither of the molecules to be detected.   Additionally, by-eye inspection of the data shows that many of the sources with only some molecular detections display evidence for additional molecules at lower S/N; therefore, it's likely that the various molecular emission features are even more strongly correlated.  However, the CO$_2$-only sources discussed in Paper I are a clear exception to this rule, and there may be others, such as Haro 1-4, which does not show strong evidence for H$_2$O, but has strong HCN and  \acetylene\ features. 

In the lower left half of Table \ref{table:detect_corr}, we show probabilities ($p$) associated with linear regression of the detections, treated as binary variables (again, excluding HAeBe and transitional disks since they are all non-detections).   For this and all future analyses, we define $p$ in a standard way, as the probability of obtaining a linear regression slope at least as high as that observed, assuming the variables are not related (i.e., the p-value associated with the t-statistic).  Thus a probability $p$ implies a correlation significant at the $(1-p)$ level.  The small $p$ for nearly all pairs of molecules shows that their detections are strongly correlated.  A notable exception is CO$_2$.

\begin{table}
\caption{Classical T Tauri detection rate and peak/continuum correlation probabilities \label{table:detect_corr}}
\begin{tabular}{ccccccc}
        &  &\multicolumn{5}{c}{Peak/Continuum} \\
&           &     H$_2$O &         OH &        HCN & C$_2$H$_2$ &     CO$_2$   \\
\multirow{4}{*}{\begin{sideways}Detections\end{sideways}} &    H$_2$O &    \ldots &      {\bf 0.00} &     {\bf  0.00} &       0.74 &      0.14  \\
 &  OH &       {\bf 0.00} &    \ldots &       {\bf 0.05} &    0.20 &       0.20  \\
   & HCN &      {\bf 0.00} &       {\bf 0.00} &    \ldots &       0.18 &       0.07  \\
    &C$_2$H$_2$ &     {\bf  0.01} &      {\bf 0.02} &   {\bf  0.00} &    \ldots &     {\bf 0.04} \\
  & CO$_2$ &       0.12 &       0.14 &       0.11 &       0.07 &    \ldots  \\
    \end{tabular}\\ \\
NOTE --- At lower left is the probability, $p$, associated with a linear regression of the two binary variables representing a detection or non-detection of the molecule. At upper right is the probability, $p$, associated with a linear regression of the peak/continuum ratios for each molecular pair.   HAeBe and transitional disks were excluded.  Bolded values have $p\leq0.05$ and are statistically significant.
\end{table}

\begin{figure}
\epsscale{1}
\plotone{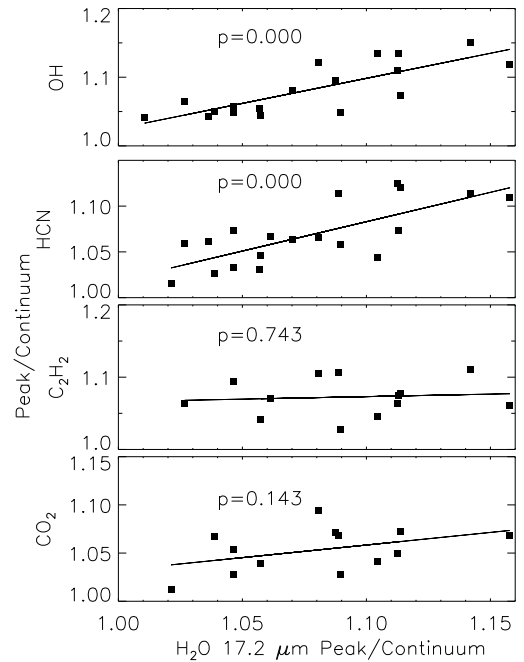}
\caption{Peak/continuum correlations for OH (27.6 $\mu$m feature), HCN, \acetylene\ and CO$_2$ with \water\ (17.2 $\mu$m feature), along with best linear fits and associated probabilities.
 \label{fig:linecont_plot}}
\end{figure}

\subsection{Line flux and peak/continuum ratio correlations between molecules}
Line fluxes and peak/continuum ratios (a ratio of the sum of the continuum and maximum line flux density to the continuum flux density)
are also strongly correlated for several of the molecules.  In Figure \ref{fig:linecont_plot}, we show the OH, HCN, \acetylene\ and CO$_2$ peak/continuum ratios against the \water\ 17.2$\,\mu$m peak/continuum ratio, along with best-fit regression models and associated probabilities, and in Table \ref{table:detect_corr} we show $p$ for all molecular pairs.   We see that peak/continuum ratios are correlated between several of the molecules, especially \water\, OH and HCN.  Line fluxes (not normalized by the continuum) are correlated for all molecular pairs, but this is largely due to the fact that all line fluxes are correlated with continuum fluxes, as we will show.  The fact that peak/continuum ratios are correlated for many molecules may mean that the line excitation is significantly influenced by qualities of the disk and radiation environment.  

\begin{figure*}
\epsscale{1}
\plotone{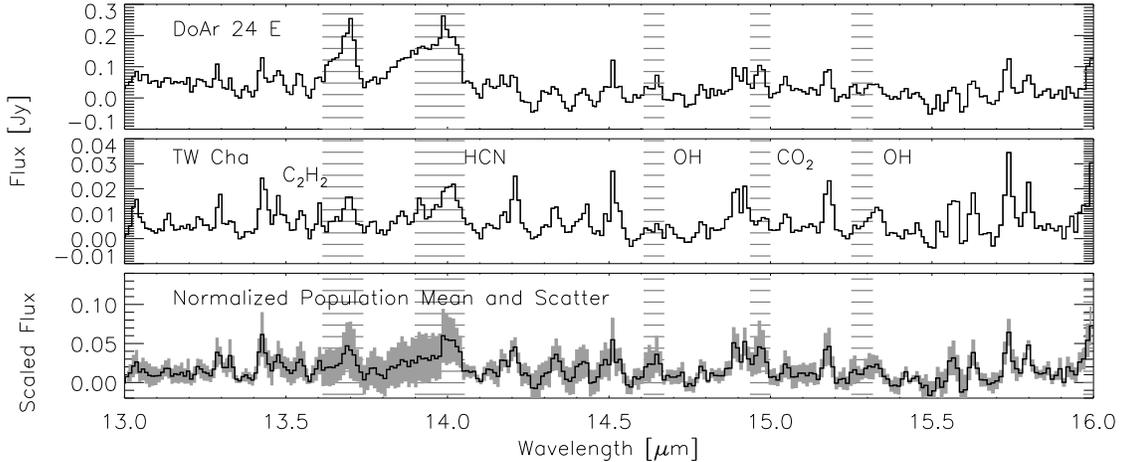}
\caption{Top two panels show 13--16 $\mu$m spectra for two sources with significantly different molecular line flux ratios.  Unlabeled features are primarily due to \water.  Below, we show mean (in black) and plus or minus one standard deviation (in gray) of the entire set of normalized spectra with observed \water\ emission.  Hatched regions from left to right mark \acetylene, HCN, OH, CO$_2$ and OH, respectively.  Note the increase in population scatter at the locations of the HCN and \acetylene\ lines.
 \label{fig:outlier_plot}}
\end{figure*}

Despite the correlations between peak/continuum ratios, variations in line ratios are still observed (and are evident as the scatter in Figure \ref{fig:linecont_plot}).  In Figure \ref{fig:outlier_plot}, for example, we show two sources --- DoAr 24E,  in which the HCN and \acetylene\ have peak/continuum ratios at least twice that of neighboring \water\ lines, and TW Cha, in which \water\ features are comparable in strength to HCN and \acetylene.  Figure \ref{fig:outlier_plot} also shows a mean and scatter of the full sample (after normalization by the 17.2 $\mu$m line).  Some features, including \water\ and OH, appear to have little cross-sample variation, while HCN and \acetylene\ show significant variation relative to \water.  

\subsection{Correlations between detections and system parameters}
In order to better understand why some sources show molecular emission and others do not, we searched for correlations between detections/non-detections and a number of stellar and disk parameters derived from the data or from the literature, including spectral type, X-ray luminosity, stellar mass, stellar luminosity, H$\alpha$ equivalent width, accretion rate, disk mass, 13 $\mu$m flux (F$_{13}$) and SED slope ($n_{13-30}$).  The stellar and disk parameters used in this analysis are shown in Tables \ref{table:star_refs} and  \ref{table:herbig}, and following Paper I, we consider the following disks to be transitional: CoKu Tau/4, HD 135344 B, SR 21, T Cha and LkH$\alpha$ 330.  Spectral types and distances can be found in Paper I.  

\begin{deluxetable*}{ccccccccccc}
\setlength{\tabcolsep}{0.02in} 
\tablecaption{T Tauri  System Characteristics \label{table:star_refs}}
\startdata
\tabletypesize{\scriptsize}
Name & EW(H$\alpha$) & $L_X$ & $\dot{M}$ & $M_\star$ &  $M_\mathrm{disk}$ & $i_\mathrm{disk}$ & $L_\star$ &Refs & F$_{13}$ & $n_{13-30}$\\
&[\AA]&[$10^{30}$ erg s$^{-1}$]&[log(M$_\odot$ yr$^{-1}$)]&[M$_\odot$]&[M$_\odot$]&[$^\circ$]&[L$_\odot$] &&[Jy]\\
\hline \\
      LkHa 270 &              92 &         \nodata  &        \nodata  &        \nodata  &        \nodata  &        \nodata  &        \nodata &                                                                 \ref{Fe95}  & $0.415\pm0.003$ & $-0.17\pm0.01$\\
       LkHa 271 &         \nodata &         \nodata  &        \nodata  &        \nodata  &        \nodata  &        \nodata  &        \nodata &                                                                          &$0.083\pm0.002$ & $0.25\pm0.04$ \\
       LkHa 326 &         \nodata &         \nodata  &        \nodata  &        \nodata  &        \nodata  &        \nodata  &        \nodata &                               &$0.340\pm0.001$&$0.171\pm0.007$                                            \\
       LkHa 327 &         \nodata &         \nodata  &        \nodata  &        \nodata  &        \nodata  &        \nodata  &        \nodata &      &$0.828\pm0.009$ & $-0.61\pm0.02$                                                                      \\
       LkHa 330 &          11-- 20 &         \nodata  &    -8.80  &        \nodata  &         0.0170  &             40  &          16.00 &                         \ref{Br07},\ref{Br08},\ref{Fe95} & $0.519\pm0.002$ & $1.933\pm0.005$\\
         LkCa 8 &         \nodata &         \nodata  &        \nodata  &        \nodata  &        \nodata  &        \nodata  &        \nodata &                         & $0.166\pm0.002$ & $0.19\pm0.02$                                                   \\
         IQ Tau &               8 &            0.42  &    $<$-8.32/-7.55  &           0.54  &         0.02  &             79  &           0.88 &  \ref{An05},\ref{Be90},\ref{Gu07a},\ref{Ki02} & $0.384\pm0.003$ & $-0.22\pm0.02$ \\
       V710 Tau &          34- 89 &            1.38  &        \nodata  &           0.42  &         0.007  &        \nodata  &           1.10 &                       \ref{An05},\ref{Gu07a},\ref{Wa09} &$0.190\pm0.002$ &$ -0.14\pm0.02$  \\
         AA Tau &              37 &            1.24  &    -8.48/-8.19  &           0.67  &         0.03  &             75  &           0.98 & \ref{An05},\ref{An07a},\ref{Be90},\ref{Gu07a},\ref{Ki02} & $0.319\pm0.002$ & $-0.32\pm0.01$ \\
     CoKu Tau/4 &           1--2 &        $< 0.21$  &      $<-11.00$  &           0.43  &         0.0005  &        \nodata  &           0.68 &               \ref{An05},\ref{Da05},\ref{Ke06},\ref{Ne95},\ref{Wa09} &  $0.119\pm0.003$ & $2.01\pm0.03$\\
         DN Tau &         \nodata &         \nodata  &    -8.46  &           0.62  &         0.03  &             30  &           1.12 &                        \ref{An05},\ref{Be90},\ref{Ha98},\ref{Is09} & $0.321\pm0.002$ &$-0.07\pm0.01$ \\
         FX Tau &          10--15 &            0.50  &    -8.65  &           0.44  &         0.0009  &        \nodata  &           1.02 &          \ref{An05},\ref{Be90},\ref{Gu07a} & $0.277\pm0.003$ & $-0.08\pm0.02$\\
         DR Tau &              87 &        $< 0.64$  &    -6.50/-6.25  &           0.55  &         0.01  &             37  &           2.50 &  \ref{An05},\ref{An07a},\ref{Be90},\ref{Is09},\ref{Ne95} & $1.957\pm0.006$ & $-0.273\pm0.005$ \\
         SX Cha &              26 &        $< 0.32$  &    -8.37  &           0.47  &        \nodata  &        \nodata  &           0.44 &                         \ref{Fe93},\ref{Ha98} & $0.508\pm0.002$ & $-0.451\pm0.007$\\
         SY Cha &          24- 64 &            0.16  &    -8.60  &           0.50  &        \nodata  &        \nodata  &           0.37 &                         \ref{Fe93},\ref{Ha98} & $0.189\pm0.001$ & $-0.150\pm0.001$ \\
         TW Cha &              26 &            1.00  &        \nodata  &           1.00  &        \nodata  &             \nodata &           0.90 &                         \ref{Fe93} &  $0.120\pm0.001$ & $-0.10\pm0.01$\\
         VW Cha &          60-147 &            9.33  &    -6.95  &           0.60  &        \nodata  &        \nodata  &           2.34 &                         \ref{Fe93},\ref{Ha98} & $0.759\pm0.002$ & $-0.103\pm0.005$\\
         VZ Cha &          58--71 &            1.26  &    -8.28  &           0.78  &        \nodata  &        \nodata  &           0.46 &                         \ref{Fe93},\ref{Ha98} & $0.389\pm0.002$ & $-1.05\pm0.01$\\
         WX Cha &              65 &            1.48  &    -8.47  &           0.54  &        \nodata  &        \nodata  &           0.68 &                         \ref{Fe93},\ref{Ha98}& $0.290\pm0.001$ & $-0.98\pm0.01$ \\
         XX Cha &             133 &        $<28.60$  &    -9.07  &           0.45  &        \nodata  &        \nodata  &           0.26 &                         \ref{Fe93},\ref{Ha98},\ref{Ro07}&$0.138\pm0.001$ & $-0.26\pm0.01$\\
          T Cha &           2--10 &         \nodata  &        \nodata  &        \nodata  &        \nodata  &        \nodata  &           1.71 &                                                     \ref{Gr05},\ref{Ke06}&$0.293\pm0.003$ &$1.60\pm0.01$\\
         Sz 50 &              46 &            3.02  &  -10.30  &        \nodata  &        \nodata  &        \nodata  &        \nodata &                                           \ref{Al00},\ref{Ke06},\ref{Sp08}&$0.208\pm0.001$&$-0.03\pm0.01$\\
     HD 135344 B &              17 &         \nodata  &    -8.30  &           1.50  &        \nodata  &             14  &           8.00 &                   \ref{Ga06},\ref{Ke06},\ref{Po08}&$0.825\pm0.003$&$1.934\pm0.005$\\
         HT Lup &           3--7 &         \nodata  &        \nodata  &           1.57  &        \nodata  &        \nodata  &           1.45 &                                           \ref{Hu94},\ref{Ke06}&$2.27\pm0.01$&$-0.33\pm0.01$\\
         GW Lup &          90--98 &         \nodata  &        \nodata  &           0.46  &        \nodata  &        \nodata  &           0.23 &                                           \ref{Hu94},\ref{Ke06}&$0.190\pm0.002$&$-0.05\pm0.02$\\
         GQ Lup &          31--39 &         \nodata  &    -8.00/-7.00  &           0.80  &        \nodata  &             51  &           0.80 &                         \ref{Ke06},\ref{Se08}&$0.507\pm0.002$&$-0.150\pm0.007$\\
         IM Lup &           4--8 &         \nodata  &        \nodata  &           1.00  &         0.1  &             50  &           1.30 &                         \ref{Hu94},\ref{Ke06},\ref{Pi08}& $0.499\pm0.002$&$-0.290\pm0.007$\\
      HD 142527 &         \nodata &         \nodata  &    -7.16  &           3.50  &        \nodata  &             30  &          69.00 &                                          \ref{Fu06}, \ref{Ga06}&$6.39\pm0.02$&$1.250\pm0.004$ \\
         RU Lup &         136-216 &         \nodata  &    -7.70  &           1.00  &         0.01 &        \nodata  &           0.42 &                       \ref{He08},\ref{Ke06},\ref{Lo09}&$4.361\pm0.004$&$-0.537\pm0.002$ \\
         RY Lup &               7 &         \nodata  &        \nodata  &           1.19  &        \nodata  &        \nodata  &           1.30 &                                             \ref{Hu94},\ref{Ke06} &$0.832\pm0.006$&$0.82\pm0.01$\\
         EX Lup &          31--43 &         \nodata  &        \nodata  &           0.59  &        \nodata  &        \nodata  &           0.39 &                                           \ref{Hu94},\ref{Ke06}&$3.173\pm0.004$&$-0.757\pm0.002$ \\
         AS 205 &          55-155 &         \nodata  &    -7.10/-6.14  &           1.50  &         0.029  &             25  &           7.10 &     \ref{An07a},\ref{An09},\ref{Fe95},\ref{Ke06},\ref{Pr03}&$7.20\pm0.02$&$-0.280\pm0.004$ \\
       Haro 1-1 &         \nodata &         \nodata  &        \nodata  &        \nodata  &        \nodata  &        \nodata  &        \nodata &                                                                      \nodata&$0.158\pm0.002$&$0.95\pm0.02$ \\
       Haro 1-4 &         \nodata &         \nodata  &        \nodata  &        \nodata  &        \nodata  &        \nodata  &        \nodata &                                                                        \nodata&$0.449\pm0.004$&$0.36\pm0.01$ \\
          VSSG1 &         \nodata &         \nodata  &    -7.19&           0.52  &         0.029  &             53  &           1.50 &                       \ref{An09},\ref{Na06}&$0.987\pm0.006$&$-0.18\pm0.01$ \\
       DoAr 24E &         \nodata &            0.50  &    -8.46 &           0.47  &         0.008  &        \nodata  &           1.26 &                        \ref{An07b},\ref{Ca95},\ref{Na06}&$2.598\pm0.004$&$-0.410\pm0.004$\\
        DoAr 25 &               2 &            0.25  &    -8.52 &           0.49  &         0.1360  &             59  &           1.35 &     \ref{An09},\ref{Ba03},\ref{Ca95},\ref{Na06}&$0.254\pm0.002$&$0.71\pm0.01$\\
          SR 21 &         \nodata &         \nodata  &        $<$-8.84  &        \nodata  &         0.005  &             22  &          24.00 &                                             \ref{An09},\ref{Br07}&$2.392\pm0.008$&$2.056\pm0.004$\\
           SR 9 &           6--14 &            2.51  &    -8.26  &           0.60  &         0.0019  &        \nodata  &           1.90 &              \ref{An07b},\ref{Ca95},\ref{Ke06},\ref{Na06}&$0.539\pm0.002$&$0.27\pm0.01$\\
       V853 Oph &              41 &         \nodata  &    -8.31 &           0.42  &        \nodata  &        \nodata  &           1.05 &                                   \ref{Na06},\ref{Sh98}&$0.441\pm0.001$&$-0.17\pm0.01$\\
        ROX 42C &         \nodata &         \nodata  &        \nodata  &        \nodata  &        $<$0.005  &        \nodata  &        \nodata &                                                                 \ref{An94}&$0.435\pm0.004$&$-0.02\pm0.02$\\
       ROX 43 A &         \nodata &         \nodata  &        \nodata  &        \nodata  &        $<$0.005  &        \nodata  &        \nodata &                                                                 \ref{An94}&$1.986\pm0.008$&$-0.306\pm0.006$\\
      Haro 1-16 &          59--76 &         \nodata  &        \nodata  &        \nodata  &         0.017  &             45  &           2.00 &                                   \ref{An09},\ref{Ke06},\ref{Sc06}&$0.539\pm0.002$&$0.850\pm0.005$\\
      Haro 1-17 &              15 &         \nodata  &        \nodata  &        \nodata  &        \nodata  &        \nodata  &        \nodata &                                                                 \ref{Ke06}&$0.155\pm0.002$&$0.28\pm0.02$\\
         RNO 90 &              76 &         \nodata  &        \nodata  &       0.9  &         0.0047  &        37  &        \nodata &                                                \ref{An07b},\ref{Ke06},\ref{Po11}&$2.162\pm0.007$&$-0.451\pm0.006$\\
       Wa Oph 6 &              35 &         \nodata  &    -6.64  &        \nodata  &         0.077 &             39  &           0.67 &                       \ref{An07a},\ref{An09},\ref{Gr05},\ref{Sh98}&$0.842\pm0.002$&$-0.386\pm0.005$\\
      V1121 Oph &         \nodata &         \nodata  &    -7.05 &        \nodata  &         0.028  &             38  &        \nodata &                                             \ref{An09} & $3.411\pm0.009$ & $-0.289\pm0.005$\\
          EC 82 &           5--11 &         \nodata  &        \nodata  &        \nodata  &        \nodata  &        \nodata  &        \nodata &                                                                 \ref{Ke06} &$1.394\pm0.002$&$0.475\pm0.003$ 
      \enddata
\tablerefs{
\usecounter{minirefcount}
\miniref{Alcal{\'a}  et al. 2000}{Al00}f
\miniref{Andre \& Montmerle 1994}{An94}
\miniref{Andrews \& Williams 2005}{An05}
\miniref{Andrews \& Williams 2007a}{An07a}
\miniref{Andrews \& Williams 2007b}{An07b}
\miniref{Andrews \& Williams 2009}{An09}
\miniref{Bary, Weintraub \& Kastner 2003}{Ba03}
\miniref{Beckwith et al. 1990}{Be90}
\miniref{Brown et al. 2007}{Br07}
\miniref{Brown et al. 2008}{Br08}
\miniref{Casanova et al. 1995}{Ca95}
\miniref{D'Alessio et al. 2005}{Da05}
\miniref{Feigelson et al. 1993}{Fe93}
\miniref{Fernandez et al. 1995}{Fe95}
\miniref{Furlan et al. 2006}{Fu06}
\miniref{Garcia Lopez et al. 2006}{Ga06}
\miniref{Gras-Vel{\'a}zquez, {\`A}., \& Ray}{Gr05}
\miniref{G{\"u}del \& Telleschi 2007}{Gu07a}
\miniref{Hartmann et al. 1998}{Ha98}
\miniref{Herczeg \& Hillenbrand 2008}{He08}
\miniref{Hughes et al. 1994}{Hu94}
\miniref{Isella, Carpenter \& Sargent 2009}{Is09}
\miniref{Kessler-Silacci et al. 2006}{Ke06}
\miniref{Kitamura et al. 2002}{Ki02}
\miniref{Lommen et al. 2009}{Lo09}
\miniref{Montesinos et al. 2009}{Mo09}
\miniref{Natta et al. 2006}{Na06}
\miniref{Neuhauser et al. 1995}{Ne95}
\miniref{Pinte et al. 2008}{Pi08}
\miniref{Pontoppidan et al. 2008}{Po08}
\miniref{Pontoppidan et al., in prep}{Po11}
\miniref{Prato, Greene \& Simon 2003}{Pr03}
\miniref{Robrade \& Schmitt 2007}{Ro07}
\miniref{Schegerer et al. 2006}{Sc06}
\miniref{Seperuelo Duarte et al. 2008}{Se08}
\miniref{Shevchenko \& Herbst 1998}{Sh98}
\miniref{Spezzi et al. 2008}{Sp08}
\miniref{Watson et al. 2009}{Wa09}
}
\tablecomments{Error bars for F$_{30}$ and $n_{13-30}$ are statistical errors.  See the text for details.}
\end{deluxetable*}

\begin{deluxetable}{ccc}
\tablecaption{HAeBe Continuum fluxes and colors \label{table:herbig}}
\startdata
Name &F$_{13}$ [Jy] & $n_{13-30}$\\
\hline \\
        HD 36112   &$2.95\pm0.01$  & $0.905\pm0.007$  \\
       HD 244604   &$1.12\pm0.01$ & $-0.37\pm0.01$  \\
        HD 36917   &$1.09\pm0.01$   &  $0.34\pm0.01$  \\
        HD 37258   &$0.95\pm0.01$ &    $-0.91\pm0.02$  \\
          BF Ori   &$0.890\pm0.006$  &  $-0.97\pm0.02$  \\
        HD 37357&   $1.15\pm0.01$  &  $-0.16\pm0.01$ \\
        HD 37411   &$0.663\pm0.003$  &  $0.875\pm0.008$ \\
          RR Tau   &$1.242\pm0.006$   &$-0.20\pm0.01$  \\
        HD 37806  & $7.30\pm0.02$   & $-1.034\pm0.006$  \\
        HD 38087   &$0.064\pm0.003$    &  $2.34\pm$0.05 \\
        HD 38120   &$6.22\pm0.02$   &  $0.243\pm0.005$  \\
        HD 50138  &$50.40\pm0.05$  &  $-1.255\pm0.003$ \\
        HD 72106   &$1.55\pm0.01$   &  $-0.26\pm0.01$ \\
        HD 95881   &$5.44\pm0.02$   &  $-1.391\pm0.008$ \\
        HD 98922  &$24.46\pm0.05$   & $-1.310\pm0.004$ \\
       HD 101412  & $2.360\pm0.008$   &  $-1.007\pm0.007$ \\
       HD 144668  &$11.35\pm0.02$ & $-1.037\pm0.006$  \\
       HD 149914   &$0.139\pm0.002$   & $-0.50\pm0.03$ \\
       HD 150193   &$9.50\pm0.02$ &   $-0.534\pm0.005$ \\
          VV Ser   &$3.34\pm0.01$   & $-1.174\pm0.006$  \\
        LkH$\alpha$ 348  & $6.82\pm0.02$ &  $-1.953\pm0.008$ \\
       HD 163296  &$10.47\pm0.02$   &$-0.485\pm0.004$ \\
       HD 179218  &$16.00\pm0.03$  & $0.269\pm0.003$ \\
       HD 190073   &$4.57\pm0.02$   &$-1.34\pm0.01$  \\
        LkH$\alpha$ 224 &  $8.0\pm0.2$ &  $0.219\pm0.003$ 
 \enddata
 \tablecomments{See Table \ref{table:star_refs}}
\end{deluxetable}

\begin{figure}
\epsscale{1}
\plotone{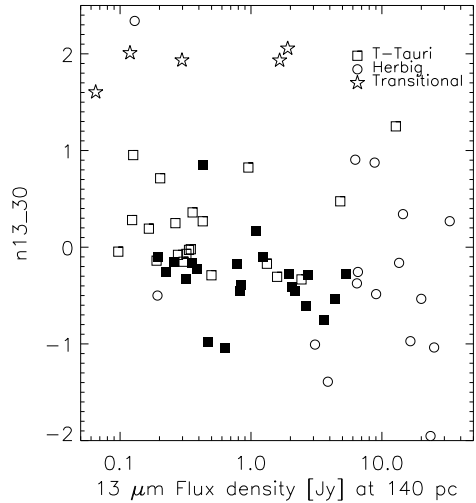}
\caption{\water\ detections and non-detections as a function of distance-normalized continuum flux and $n_{13-30}$ color.  Detected sources are shown by filled symbols.  
 \label{fig:color_plot}}
\end{figure}

In Figure \ref{fig:color_plot}, we show two major trends seen in the data --- those between \water\ detections and (distance-normalized) F$_{13}$ or $n_{13-30}$.  These trends are related to the conclusions presented in Paper I --- that HAeBe disks and transitional disks are much less likely to show emission.  For example,  there are no \water\ detections for  F$_{13}\gtrsim 6$, which represent the HAeBe disks.    
Similarly, we see that there are no \water\ detections for  $n_{13-30}\gtrsim 1$, which corresponds to non-detections in transitional disks.  However, Haro 1-16, which has detected \water\ emission, can be considered transitional by some definitions \citep{Salyk09} and shows evidence for inner clearing in millimeter maps \citep{Andrews09}, making it a ``borderline'' transitional, or perhaps pre-transitional \citep{Espaillat07}, disk.  

To look for other interesting trends in the data without being influenced by the transitional and HAeBe non-detections, we also considered a restricted sample, which included only cTTs disks.  Results are shown in Table \ref{table:detect_param_corr}. Several parameters show high statistical significance with detections; however, with 45 parameter/molecule pairs, we would expect a few pairs to be significant at the 95\% level through chance alone, so these may or may not be significant.  However, a notable result is that detections for all molecules are anti-correlated with $n_{13-30}$ at at least the 90\% level. $n_{13-30}$ is a measure of the degree to which small grains have settled out of the disk atmosphere \citep[e.g.][]{Furlan06}, with small $n_{13-30}$ corresponding to a higher degree of settling.  Therefore, the observed correlation is consistent with greater detection rates in disks in which small grains have begun to settle to the midplane.   This result follows naturally from the fact that settling of small grains allows us to see deeper into the disk, and hence observe a larger column and higher density (i.e., a density closer to the critical density) of gas.    However, the detailed (and likely complex) effects of grain settling on gas heating and structure have yet to be explored.

\begin{figure}
\epsscale{1}
\plotone{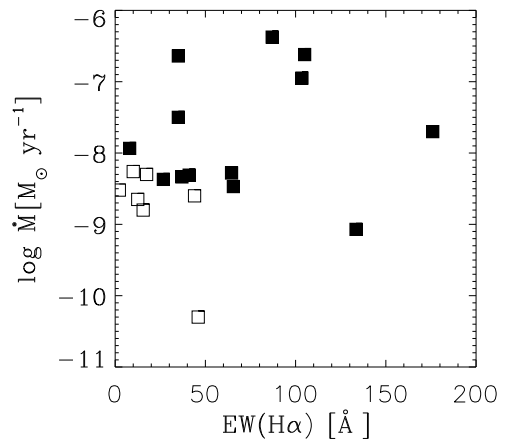}
\caption{\water\ detections and non-detections in cTTs as a function of H$\alpha$ and $\dot{M}$.  Detected sources are shown by filled symbols.  
 \label{fig:detect_mdot_plot}}
\end{figure}

We also find that all molecular detections are correlated with H$\alpha$ at the 90\% level or more, although, interestingly, only \water\ detections show a correlation with accretion rate.  
Water detections as a function of H$\alpha$ and $\dot{M}$ are shown graphically in Figure \ref{fig:detect_mdot_plot}.  The correlation with $\dot{M}$ must be approached with some caution, as the $\dot{M}$'s derive from a variety of methods, which are known to have systematic offsets and significant scatter \citep[e.g.][]{Herczeg08}. In fact, since higher $\dot{M}$'s tend to be derived from UV excess measurements, while lower $\dot{M}$'s are derived from line observations, we find that reducing the UV-excess-derived values by factors of four to seven (the offset between UV-excess and H$\alpha$ measurements found by Herczeg et al. 2008) can render the $\dot{M}$ detection trend statistically insignificant.  However, the trend with H$\alpha$ remains significant, and could mean that accretional heating is a necessary ingredient for the excitation of mid-IR molecular emission.  

\begin{deluxetable*}{lrrrrrrrrr}
\tablecaption{Regression probabilities for detections vs. cTTs parameters \label{table:detect_param_corr}}
\startdata
          Molecule & EW(H$\alpha$) &      L$_X$ &  $\log \dot{M}$ &  $M_\star$ &      $M_d$ &        $i$ &  $L_\star$ &        F13 &    n13\_30   \\
       H$_2$O &    \bf{ 0.00} &          0.54 &    \bf{ 0.02} &      (-) 0.21 &      (-) 0.58 &          0.21 &      (-) 0.11 &          0.16 & (-)\bf{ 0.00}  \\
           OH &    \bf{ 0.04} &          0.36 &          0.23 &      (-) 0.30 &      (-) 0.79 &          0.82 &      (-) 0.20 &          0.58 & (-)\bf{ 0.00}  \\
          HCN &    \bf{ 0.02} &      (-) 0.55 &          0.28 & (-)\bf{ 0.03} &          0.86 &    \bf{ 0.05} & (-)\bf{ 0.05} &      (-) 0.90 &      (-) 0.06  \\
   C$_2$H$_2$ &          0.09 &      (-) 0.86 &          0.65 &      (-) 0.10 &          0.80 &    \bf{ 0.04} &      (-) 0.21 &      (-) 0.97 &      (-) 0.07  \\
       CO$_2$ &    \bf{ 0.00} &          0.30 &          0.53 &      (-) 0.50 &          0.77 &          0.85 &      (-) 0.20 &          0.42 & (-)\bf{ 0.00}  \\
\enddata
\tablecomments{Probabilities associated with the linear regression slope.  Statistically significant correlations (p$\leq$0.05) are shown in bold. A (-) in front of the value means that it is an anti-correlation.}
\end{deluxetable*}

\subsection{Correlations between line fluxes and system parameters}
\begin{figure*}
\epsscale{1}
\plotone{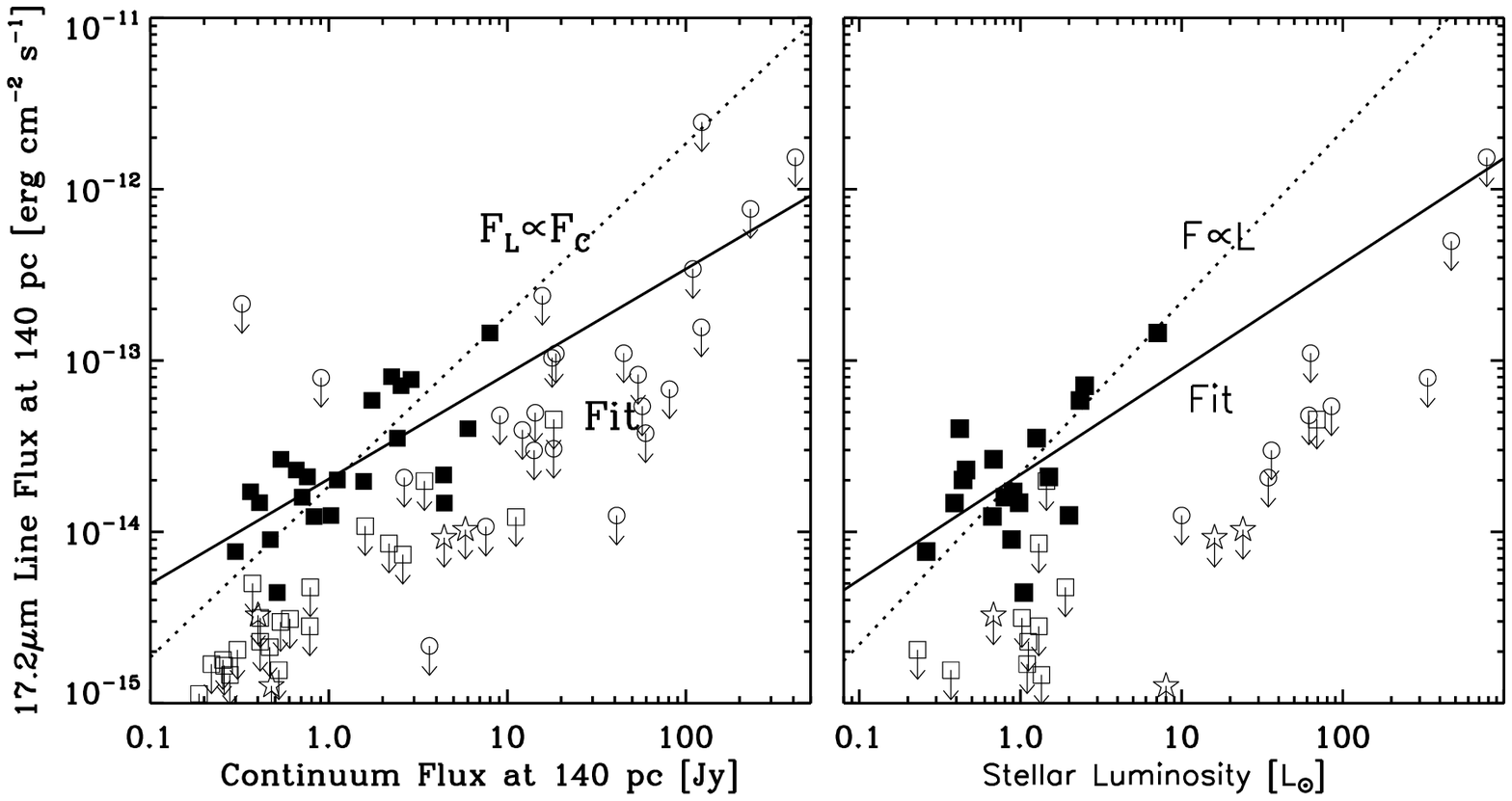}
\caption{Correlations between \water\ line flux and distance-normalized continuum flux or stellar luminosity.   Solid lines show best fits to sources with detected \water\ emission.  Dashed lines show best constant line/continuum fit and best fit with line flux proportional to stellar luminosity, respectively.  Symbols are defined in Figure \ref{fig:color_plot}.  
 \label{fig:lum_plot}}
\end{figure*}

In Table \ref{table:flux_param_corr}, we show $p$ for line fluxes versus various stellar and disk parameters.   {\it All} molecular line fluxes are correlated with the continuum flux, F$_{13}$, 
as well as with several other parameters.  In Figure \ref{fig:lum_plot}, we show the distance-normalized line flux of the 17.2 $\mu$m \water\ line against the 17 $\mu$m continuum flux.
The dashed line has slope fixed at 1, which is equivalent to assuming that the line flux varies in a sense proportional to continuum flux.  We find that the \water\ line flux is correlated with continuum flux, although the actual best-fit slope, $0.61\pm0.14$, differs somewhat from direct proportionality.  Additionally, there is significant scatter about the trend, showing an intrinsic range in the line/continuum ratios
among detected sources.  Figure \ref{fig:lum_plot} also shows that HAeBe disks have line fluxes lower than the trend predicted by the cTTs emission lines, consistent with the conclusions of Paper I.

\begin{deluxetable*}{lrrrrrrrrr}
\tablecaption{Regression probabilities for line fluxes vs. cTTs parameters \label{table:flux_param_corr}}
\startdata
       Molecule     & EW(H$\alpha$) &      L$_X$ &  $\dot{M}$ &  $M_\star$ &      $M_d$ &        $i$ &  $L_\star$ &        F13 &    n13\_30   \\
      H$_2$O (15.2 $\mu$m) &       0.06 & \bf{ 0.04} & \bf{ 0.03} & \bf{ 0.00} &   (-) 0.56 & \bf{-0.04} & \bf{ 0.00} & \bf{ 0.00} &   (-) 0.63  \\
      H$_2$O (17.2 $\mu$m) &       0.08 & \bf{ 0.01} & \bf{ 0.02} & \bf{ 0.00} &   (-) 0.46 & \bf{-0.05} & \bf{ 0.00} & \bf{ 0.00} &   (-) 0.58  \\
          OH (23.2 $\mu$m) &       0.55 & \bf{ 0.00} & \bf{ 0.05} &       0.22 &       0.81 &   (-) 0.11 & \bf{ 0.05} & \bf{ 0.00} &   (-) 0.60  \\
          OH (27.6 $\mu$m) & \bf{ 0.04} &       0.12 & \bf{ 0.05} & \bf{ 0.02} &   (-) 0.46 &   (-) 0.07 & \bf{ 0.00} & \bf{ 0.00} &       0.97  \\
         HCN &       0.28 &       0.92 &       0.18 &       0.21 &   (-) 0.32 &   (-) 0.21 & \bf{ 0.02} & \bf{ 0.03} &   (-) 0.20  \\
  C$_2$H$_2$ &       0.10 &   (-) 0.83 &       0.37 &       0.29 &   (-) 0.29 &   (-) 0.12 &       0.26 & \bf{ 0.00} &   (-) 0.23  \\
      CO$_2$ &       0.56 & \bf{ 0.03} &       0.18 & \bf{ 0.01} &   (-) 0.70 &   (-) 0.19 & \bf{ 0.00} & \bf{ 0.00} &   (-) 0.53  \\
\enddata
\tablecomments{Same notation as Table  \ref{table:detect_param_corr}.}
\end{deluxetable*}

As can be seen in Table \ref{table:flux_param_corr}, the correlation between continuum and line flux implies a correlation with several other system parameters that covary with the continuum flux.  One such correlation, between line flux and $L_\star$, is shown in the second panel of Figure \ref{fig:lum_plot}, and is marginally consistent with F$\propto$L.  This could be related to increased heating of the gas, or an increase in the solid angle of the emitting region (due to gas being heated at larger distances from the star).   In fact, such a relationship between luminosity and solid angle has been found for CO (Pontoppidan et al.\ 2011; C. Salyk et al., in preparation).  The significance of the correlation observed here may, however, depend on a few stars with high mass accretion rates.  

To look for other correlations with disk or stellar parameters that are {\it independent} of the line-to-continuum correlation, we use peak/continuum ratios, thereby accounting for and removing the line/continuum correlation.  However, we find that virtually all significant correlations with line fluxes are removed once we account for the continuum flux correlation.  In fact, the number of parameters with significant correlations is approximately what would be expected by chance alone.

\subsection{CO detections and correlations}
\begin{figure}
\epsscale{1}
\plotone{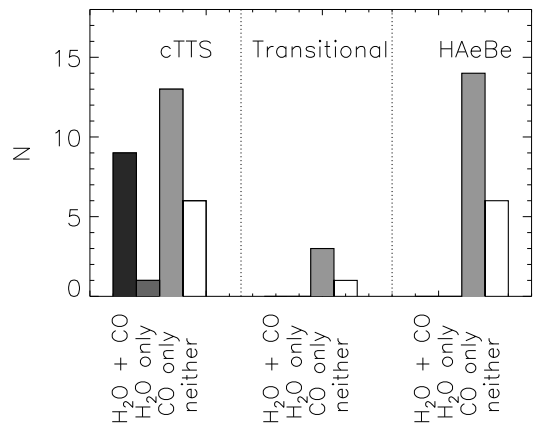}
\caption{Histogram showing detections of mid-IR \water\ and vibrational CO emission.   \label{fig:co_hist}}
\end{figure}

A majority of the disks in our sample were observed in the M band with Keck-NIRSPEC and most of these show CO rovibrational emission --- including most HAeBe disks and many transitional disks. Figure \ref{fig:co_hist} shows a histogram of detections and non-detections comparing CO and \water.  The majority of both HAeBe and transitional disks show CO in emission, while neither show strong mid-IR molecular emission.  (Again, it must be mentioned that Haro 1-16 has some characteristics of a transitional or pre-transitional disk, and has evidence for \water\ emission).

\begin{figure}
\epsscale{0.8}
\plotone{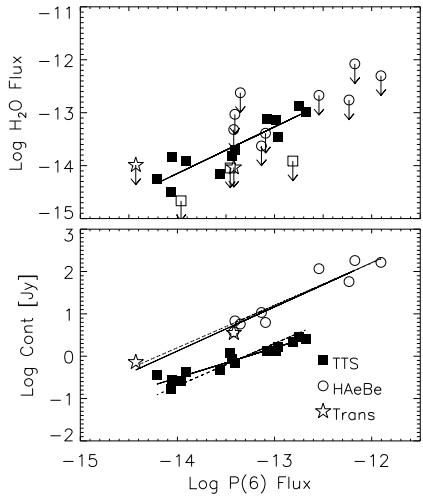}
\caption{Correlations between CO P(6) and 17.2 $\mu$m \water\ line fluxes (in erg cm$^{-2}$ s$^{-1}$) and M-band continuum flux density (in Jy), all normalized to 140 pc. Solid lines are best least-squares fits, while dashed lines are best constant line/continuum fits.  Arrows denote 3$\sigma$ upper limits.  Filled symbols are sources with detected water, while open symbols are non-detections.
 \label{fig:co_corr}}
\end{figure}

In Figure \ref{fig:co_corr}, we show that distance-normalized CO and \water\ line fluxes are highly correlated (with $p$=0.004).  Interestingly,  for most HAeBe disks, the observed \water\ upper limits are consistent with what would be predicted from the CO line fluxes.  Thus, the data are at least consistent with no difference in CO/\water\ line ratios between cTTs and HAeBe disks, to within the constraints imposed by the upper limits on the \water\ line flux.  Instead, as shown in Figure \ref{fig:co_corr}, we find that CO line-to-continuum ratios are systematically lower for HAeBe and transitional disks, just as was found for \water\ (Paper I), making the CO/\water\ ratios similar for all classes of objects.  

\section{LTE Models}
\label{sec:lte}
\subsection{Description of modeling procedures}
\subsubsection{General modeling procedure}
To obtain a better physical understanding of the source of the disk emission, we fit the IRS data with 
emission models.  In this work, we use an isothermal slab approximation, in which the emitting area $A$, temperature $T$, and column density $N$ are all free parameters.  These models are the same as those utilized in \citet{Salyk08}, and assume a Gaussian local line shape with width $\sigma$=2 km s$^{-1}$.  We make no correction for disk inclination (since this has been measured for only a small subset of our sample), so the $A$ we derive is a projected area, and $N$ is a slant column density.  

The lines are not spectrally resolved with the Spitzer-IRS, and so the global line shape (which depends on large-scale, largely Keplerian motion, and disk orientation) as well as the local line shape (which depends on the local gas velocity) are both unknown, but only the local line shape affects the slab model line flux.  We choose a moderate value of the local line width, $\sigma$, equivalent to the thermal velocity of H$_2$ at $\sim$700 K, or $\sim$0.07$\,v_\mathrm{Kep}$ at 1 AU around a solar-mass star.  This is a few times larger than the thermal velocity of H$_2$O at 600 K and thus assumes some non-thermal gas motion.  Since the actual local line broadening is essentially unknown, we consider here the effect of our chosen assumption.  For molecules like \water\, which have emission features at a variety of optical depths ($\tau$), the best-fit $N$ scales roughly linearly with $\sigma$, since the data dictate $\tau$, and $\tau\propto N/\sigma$.  For molecules like HCN, for which the emission is likely optically thin, the choice of $\sigma$ does not affect the total observed line flux, and thus does not affect the best-fit $N$, unless $\sigma$ is so small ($\lesssim 0.1$ km s$^{-1}$), that the emission becomes optically thick.  However, such small line widths are unlikely, as velocities cannot be sub-thermal, and $T$ is several hundred K for all molecules.

There are many limitations to using this simple slab model.  The actual disk environment is quite complex, with
densities and temperatures that depend on both height and radius, so the slab model only represents some average
conditions in the disk atmosphere.  Further, it has been shown that the line-excitation is not in LTE \citep{Pontoppidan10b},
as was predicted by radiative transfer models  \citep[e.g.][]{Meijerink09}.  Many of the more complex issues related to line excitation 
and disk structure are being pursued with these sophisticated models.

On the other hand, the slab models offer the benefit of providing general characteristics of the emitting gas for a large number of disks, with a relatively small number of complications and free parameters.  Also, at least for \water, non-LTE/LTE flux ratios remain close to unity for many transitions \citep{Meijerink09}, and the LTE results appear to reproduce the observed spectra reasonably well \citep{Carr08, Salyk08}.  Here, we use the slab models to understand the bulk properties of the emitting atmospheres, to investigate the variety of emission observed, and to estimate molecular abundance ratios.  

\subsubsection{Modeling procedure for \water}
\begin{figure}
\epsscale{0.9}
\plotone{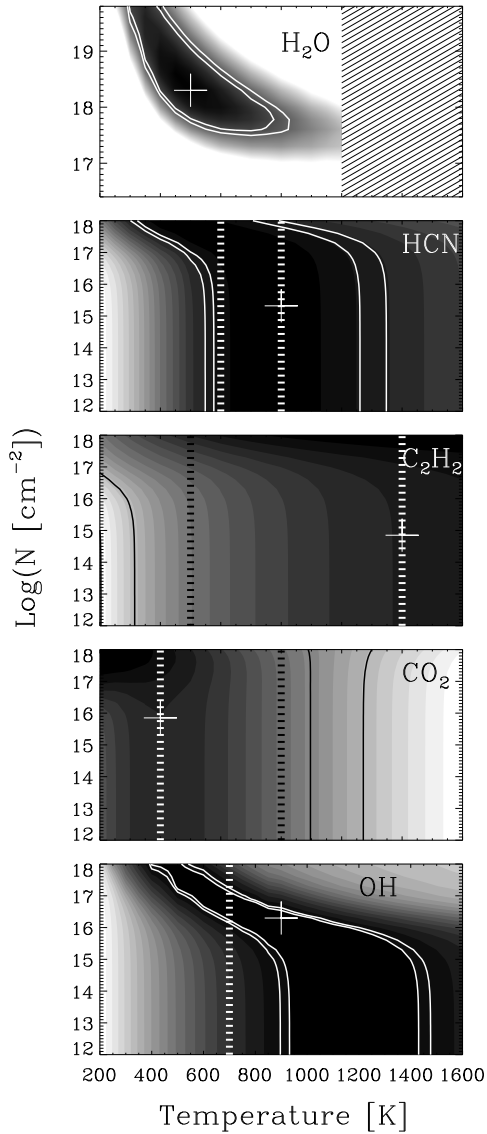}
\caption{$\chi^2$ as a function of $N$ and $T$ for VW Cha, along with  10\% and 5\% probability contours. Each individual model is scaled by its own factor ($A$) to minimize residuals, and $\sigma$ is fixed such that $\chi_\mathrm{red}^2=1$ for the best fit.  White crosses show best-fit parameters used in our analysis, where for all molecules except for \water, $A$ is fixed to $A_\mathrm{H_2O}$ and $T$ is limited to lie between the vertical dashed lines
(see also \S \ref{sec:modeling}). 
 \label{fig:chisq_plot}}
\end{figure}

The \water\ emission spectrum observed by the IRS consists of lines with a large range of excitation temperatures and opacities, and
so it is possible, in principle, to constrain all three model parameters ($A$, $T$, and $N$).  In particular, $A$ is principally determined from
the strengths of the optically thick lines,  while $N$ and $T$ both affect the relative strengths of lines as a function of wavelength.  
As discussed in \citet{Meijerink09}, there is only a small change in line strength across a wide range of wavelengths, in contrast to predictions from simple disk models.
Those authors interpret this observation as a depletion of water beyond $\sim$ 1 AU, perhaps due to a ``cold-finger'' effect.  The slab model cannot shed light on the 
radial distribution of the water abundance and the ``flatness'' of the spectrum is instead produced by a combination of high $T$ and $N$. (However, the fact that the slab model provides as good a fit as it does is consistent with the ``cold-finger'' effect hypothesis, as the spectra are reasonably well-fit by a single, relatively warm emitting temperature, equivalent to an inner disk emitting region with a sharp outer cutoff.)  In Figure \ref{fig:chisq_plot}, we show the dependence of $\chi^2$ on $N$ and $T$, allowing $A$ to be a tunable variable, and note that there is a well-defined minimum, but that $N$ and $T$ are somewhat degenerate.   

Because of the complexity of the water spectrum, and the limitations inherent to the simple model, no slab model provides a perfect fit to the data, and different fitting methods can produce different outcomes.   We chose to compute 65 line peaks for data and model, and minimize the difference between these two, with line peaks computed as the peak flux density minus an estimate of the local continuum.  The set of lines includes any feature that could be reasonably distinguished from surrounding features, and the lines span wavelengths from 10--35 $\mu$m and thus a large range in excitation.  This method does not require a global continuum fit, since it instead defines the continuum locally around each emission line.  It is also not flux-weighted, and so takes weaker short-wavelength lines into account.  We performed the fits by iterating over a grid of $N$ and $T$, choosing a best-fit $A$ for each of these pairs, and then choosing the best overall fit (based on the rms residuals).  

\subsubsection{Modeling procedure for HCN, \acetylene, CO$_2$ and OH}
\label{sec:modeling}
Continuum-subtracted spectra were used to fit all other molecular features.  The continuum was determined by fitting a spline to points along the spectrum where no strong \water\ or other emission features were expected.  The preferred method and number of continuum points was carefully chosen to allow removal of broad-band features, such as those due to solid-state silicates, but not to remove narrower features due to gas molecules.  However, uncertainties in continuum determination are unavoidable, and must be considered when determining the accuracy of any model fits. 

\begin{figure*}
\epsscale{1.}
\plotone{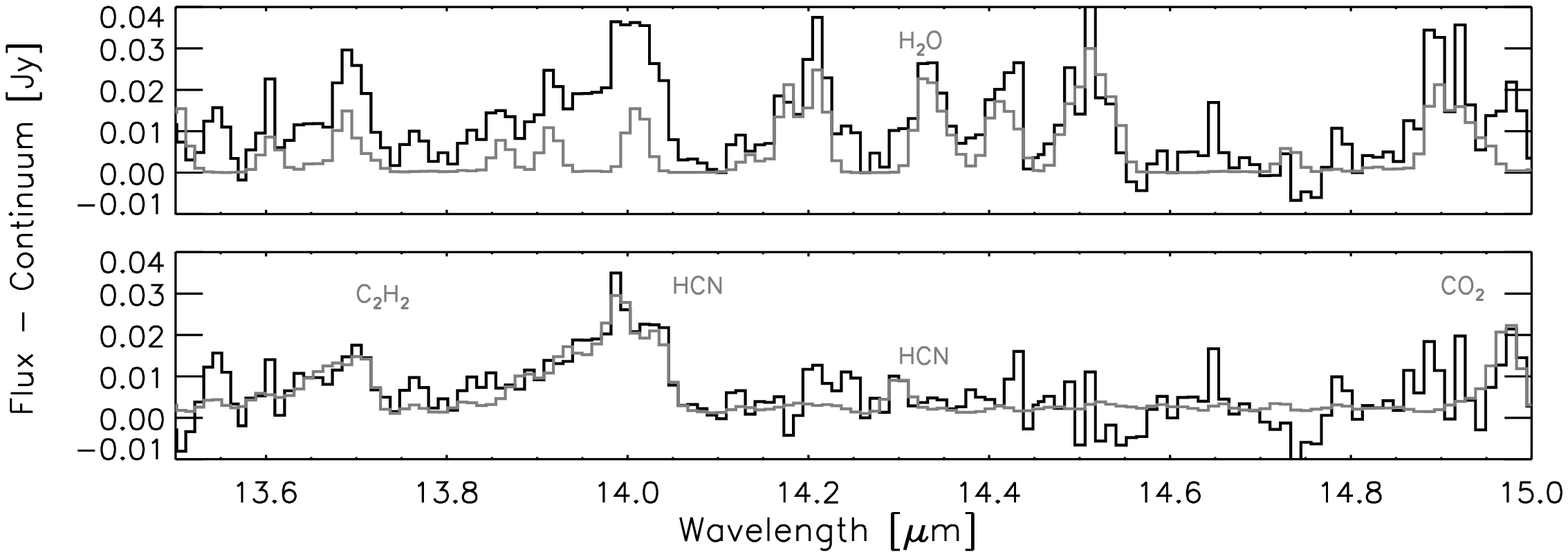}
\caption{Example (for WX Cha) of \water\ model removal and its effect on other molecular line shapes and strengths. The top panel shows the reduced data (in black) and best-fit \water\ model (in gray).  The bottom panel shows the data after \water\ model removal, along with a model that includes HCN, \acetylene\ and CO$_2$.
 \label{fig:removal_plot}}
\end{figure*}

After continuum subtraction, the best-fit water model was also subtracted.  We determined that our nominal \water\ fits, which used the entire IRS wavelength range, did not produce a good match to the data below $\sim 14 \mu$m. Therefore, we performed a separate fit to water lines in the 10--14 $\mu$m region (which typically require temperatures increased by $\sim$100 K and areas reduced by a factor of $\sim$2), and subtracted these fits below 14 $\mu$m.  Finally, it was important to remove contributions from all other molecules; we began by ignoring these contributions, and then performed a few iterations of model fits and subtractions.  Figure \ref{fig:removal_plot} shows an example of the water model subtraction and its effect, for WX Cha.  Some features are well-subtracted, but others are not, and these residuals affect the remaining molecular line shapes. 

For molecules except for water,  the models are sufficiently degenerate that $T$, $N$ and $A$ can not be determined simultaneously given our (primarily systematic) uncertainties.  This is illustrated in Figure \ref{fig:chisq_plot}, which shows $\chi^2$ as a function of $T$ and $N$, where $A$ is adjusted to minimize residuals, and $\chi^2_{red}$ is set to 1 for the best fit. At low $N$ (in the optically-thin regime) $A$ and $N$ are completely degenerate, and so $\chi^2$ is insensitive to changes in $N$.  $\chi^2$ contours curve as $N$ enters the optically-thick regime, as the line shapes begin to be affected by the opacity, but fits still remain somewhat degenerate with respect to all three fit parameters. Even when distinct minima exist (as in the case of CO$_2$) they are not significant in a statistical sense, because of uncertainties in the continuum- and water-model-subtracted data.  Therefore, we have chosen not to fit all three parameters simultaneously.

\begin{figure*}
\epsscale{1}
\plotone{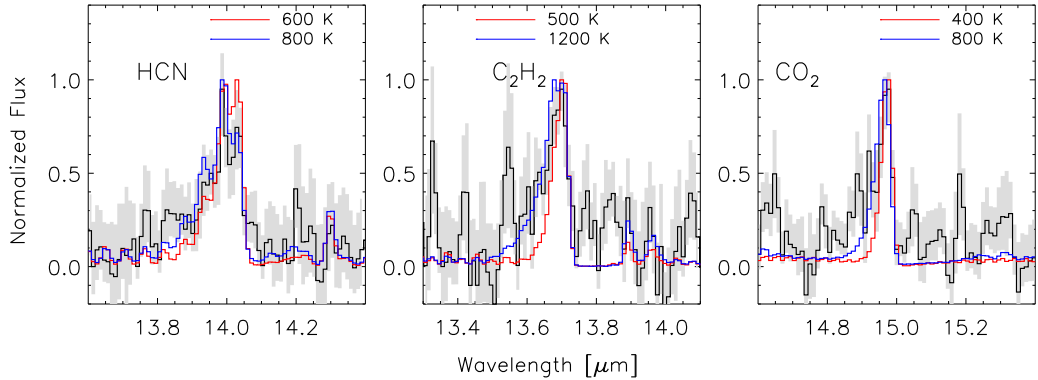}
\caption{Mean spectra, normalized to the line peak, in black.  Gray areas delineate plus or minus 1 standard deviation from the mean.  Normalized models, at temperatures meant to bracket the observed data, are shown in red and blue.
 \label{fig:shape_plot_all}}
\end{figure*}

The following procedure was chosen to fit the molecular emission features. First, an appropriate range of $T$ was determined by computing a normalized mean emission feature and comparing with models, as shown in Figure \ref{fig:shape_plot_all}.  Then, fixing $A$ to $A_\mathrm{H_2O}$, $T$ (within the allowed range) and $N$ were varied to minimize $\chi^2$ (over a wavelength range wide enough to include all potential emission).  A model with a different $A$ could potentially produce a lower $\chi^2$ (Figure \ref{fig:chisq_plot}), but we believe that the systematic uncertainties do not justify the choice of this model over others. This differs from the procedure chosen by \citep{Carr08}, who simultaneously fit all three model parameters for all molecules except \acetylene.  We stress that our modeling results are not inconsistent with theirs, but rather represent a more conservative approach based on our understanding of the systematic uncertainties.  Some care must be taken with the interpretation of our results, as the effective emitting regions are almost certainly different for each molecule.  However, column densities can easily be adjusted as additional data place constraints on $A$.   The change in $N$ would be directly anti-proportional to changes in $A$ in the optically thin regime, but changes in $N$ will be more significant if the lines turn out to be optically thick.

To compute upper limits, we first estimated the emitting area, either by setting it equal to that derived for \water\ (if observed) or by using the results of a linear fit to best-fit emitting area versus distance-normalized $F_{13}$. For each molecule, we then chose a characteristic temperature (shown as dashed lines in Figure \ref{fig:hist_plot}) and used the assumed emitting area to compute emission models.  We used a 3$\sigma$ detection limit for the line (or band) peak to compute a column density upper limit.  For \water\, we used the 17.2 $\mu$m line limits.

\subsubsection{Modeling procedure for CO}
We fit CO line fluxes for most CO-emitting sources with an LTE slab model following the procedure described in \citet{Salyk09}.   A few sources were omitted due to low S/N (HD 244604, HD 36917, RR Tau and HD 179218) and SR 21 was omitted because it is not well fit by an LTE model \citep{Salyk09}.  A rotation diagram was constructed from CO emission line fluxes, and these were fit with a grid of LTE models with $T$, $N$ and $A$ as free parameters.  In addition, we used $^{13}$CO fluxes or non-detections as a constraint on $N$.  Finally, we restricted our analysis to those sources with at least two line flux measurements at $J\, >20$, as we found the fits otherwise unreliable. 

\subsection{Model Results}

\subsubsection{\water\ Results}

\begin{figure}
\epsscale{.9}
\plotone{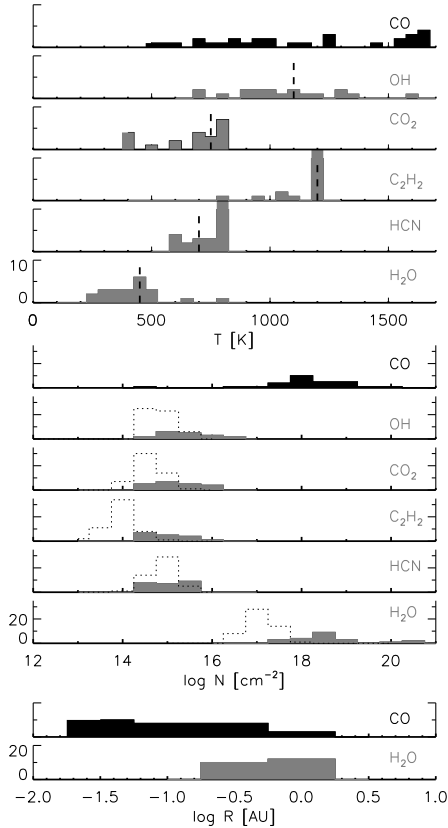}
\caption{Filled, gray histograms show best-fit model parameters from LTE fits.   $A_\mathrm{H_2O}$ is used for all molecules except for CO.  Dotted, unfilled histograms show upper limits, while dashed vertical lines show the assumed temperature used to compute the upper limits.
 \label{fig:hist_plot}}
\end{figure}

Best-fit \water\ model parameters and upper limits are shown in Tables \ref{table:moltable} and \ref{table:moltable_b}, and a histogram of these results is shown in 
Figure \ref{fig:hist_plot}.  Best-fit $T$'s  cluster around 450 K, and best-fit $N$'s cluster around 10$^{18}\, \mathrm{cm}^{-2}$.  Temperatures of 450 K correspond to blackbody dust at a distance of 1.65 AU at the surface of a passively heated disk with $T_\star=4000\ $K and $R_\star=2.5 R_\odot$  \citep{Chiang97}.  However, models predict that the gas in the disk's upper layers thermally decouples from the dust, so warm gas may exist at much larger distances from the star \citep[e.g.][]{Glassgold01, Kamp04}.  $N_\mathrm{H_2O}$ is close to that derived from the water self-shielding process proposed by \citet{Bethell09}, but it should be kept in mind that the \water\ column densities in their models are derived in part from the IRS observations themselves.  

\begin{deluxetable*}{cccccccccccc}
\tablecaption{Model Fit Results -  cTTs Systems \label{table:moltable}}
\tabletypesize{\scriptsize}
\startdata
Name & T$_\mathrm{H_2O}$ & log N$_\mathrm{H_2O}$ & T$_\mathrm{OH}$ & log  N$_\mathrm{OH}$ &  T$_\mathrm{HCN}$ & log  N$_\mathrm{HCN}$ &  T$_\mathrm{C_2H_2}$ & log N$_\mathrm{C_2H_2}$  &  T$_\mathrm{CO_2}$ &  log N$_\mathrm{CO_2}$ &R \\
       LkHa 270 &      450\tablenotemark{*}&                  $<$17.6&    1100\tablenotemark{*}&                  $<$15.3&                      700\tablenotemark{*}&                  $<$15.4&    1200\tablenotemark{*}&                  
$<$14.4&                      700&                     15.6&      1.30\\
       LkHa 271 &      450\tablenotemark{*}&                  $<$17.4&    1100\tablenotemark{*}&                  $<$15.5&                      700\tablenotemark{*}&                  $<$15.3&    1200\tablenotemark{*}&                  
$<$14.0&     750\tablenotemark{*}&                  $<$15.1&      1.02\\
       LkHa 326 &                       650&                     18.6&                      900&                     16.3&                      600&                     15.8&    1200\tablenotemark{*}&                  
$<$15.0&                      700&                     16.3&      0.45\\
       LkHa 327 &                       350&                     18.8&                     1350&                     14.9&                      700\tablenotemark{*}&                  $<$15.2&    1200\tablenotemark{*}&                  
$<$14.3&                      800&                     15.3&       2.4\\
       LkHa 330 &      450\tablenotemark{*}&                  $<$17.4&    1100\tablenotemark{*}&                  $<$15.6&                      700\tablenotemark{*}&                  $<$15.1&    1200\tablenotemark{*}&                  
$<$14.2&     750\tablenotemark{*}&                  $<$14.9&      1.37\\
         LkCa 8 &      450\tablenotemark{*}&                  $<$17.0&    1100\tablenotemark{*}&                  $<$14.9&                      700\tablenotemark{*}&                  $<$14.9&    1200\tablenotemark{*}&                  
$<$13.7&     750\tablenotemark{*}&                  $<$14.6&     0.990\\
         IQ Tau &                       800&                     17.6&    1100\tablenotemark{*}&                  $<$15.7&                      800&                     15.8&                     1200&                  
   15.6&     750\tablenotemark{*}&                  $<$15.5&      0.49\\
       V710 Tau &      450\tablenotemark{*}&                  $<$17.0&    1100\tablenotemark{*}&                  $<$14.8&                      750&                     15.0&                     1200&                  
   15.0&                      800&                     14.8&      1.00\\
         AA Tau &                       400&                     18.8&                     1000&                     15.6&                      800&                     15.6&                     1100&                  
   15.3&                      600&                     15.3&      0.74\\
     CoKu Tau/4 &      450\tablenotemark{*}&                  $<$17.3&    1100\tablenotemark{*}&                  $<$14.9&                      700\tablenotemark{*}&                  $<$15.0&    1200\tablenotemark{*}&                  
$<$13.9&     750\tablenotemark{*}&                  $<$14.9&     0.980\\
         DN Tau &      450\tablenotemark{*}&                  $<$17.1&    1100\tablenotemark{*}&                  $<$14.9&                      700\tablenotemark{*}&                  $<$14.9&    1200\tablenotemark{*}&                  
$<$13.9&     750\tablenotemark{*}&                  $<$14.7&      1.04\\
         FX Tau &      450\tablenotemark{*}&                  $<$17.2&    1100\tablenotemark{*}&                  $<$14.9&                      600&                     15.0&    1200\tablenotemark{*}&                  
$<$14.0&     750\tablenotemark{*}&                  $<$14.9&      1.02\\
         DR Tau &                       500&                     18.6&                     1050&                     15.6&                      800&                     15.6&                     1200&                  
   14.8&                      400&                     15.9&       1.1\\
         SX Cha &                       350&                     18.9&                     1300&                     15.3&                      650&                     14.7&    1200\tablenotemark{*}&                  
$<$14.0&     750\tablenotemark{*}&                  $<$14.8&       1.0\\
         SY Cha &      450\tablenotemark{*}&                  $<$16.9&    1100\tablenotemark{*}&                  $<$14.6&                      650&                     15.0&                     1200&                  
   14.9&                      800&                     15.0&      1.04\\
         TW Cha &                       450&                     18.6&                     1150&                     15.3&                      750&                     15.4&                     1200&                  
   15.2&     750\tablenotemark{*}&                  $<$14.6&      0.76\\
         VW Cha &                       500&                     18.3&                      900&                     16.3&                      800&                     15.3&                     1200&                  
   14.8&                      400&                     15.8&       1.2\\
         VZ Cha &                       450&                     18.6&                     1100&                     15.8&                      800&                     15.5&                      800&                  
   15.4&     750\tablenotemark{*}&                  $<$14.9&      0.78\\
         WX Cha &                       450&                     18.8&                      950&                     15.6&                      800&                     15.6&                     1200&                  
   14.7&                      400&                     15.7&      0.79\\
         XX Cha &                       250&                     20.3&                      700&                     16.0&                      800&                     15.0&                     1200&                  
   15.7&                      700&                     15.2&      0.88\\
          T Cha &      450\tablenotemark{*}&                  $<$16.6&    1100\tablenotemark{*}&                  $<$14.6&                      700\tablenotemark{*}&                  $<$14.4&    1200\tablenotemark{*}&                  
$<$13.4&     750\tablenotemark{*}&                  $<$14.2&     0.960\\
          Sz 50 &      450\tablenotemark{*}&                  $<$17.0&    1100\tablenotemark{*}&                  $<$14.7&                      700\tablenotemark{*}&                  $<$14.7&    1200\tablenotemark{*}&                  
$<$13.9&                      750&                     15.0&      1.04\\
     HD 135344 B &      450\tablenotemark{*}&                  $<$16.8&    1100\tablenotemark{*}&                  $<$14.8&                      700\tablenotemark{*}&                  $<$14.6&    1200\tablenotemark{*}&                  
$<$13.6&     750\tablenotemark{*}&                  $<$14.3&      1.03\\
         HT Lup &      450\tablenotemark{*}&                  $<$17.7&    1100\tablenotemark{*}&                  $<$15.0&                      700\tablenotemark{*}&                  $<$15.4&    1200\tablenotemark{*}&                  
$<$14.5&                      800&                     15.4&      1.57\\
         GW Lup &      450\tablenotemark{*}&                  $<$17.1&    1100\tablenotemark{*}&                  $<$14.9&                      700\tablenotemark{*}&                  $<$14.9&    1200\tablenotemark{*}&                  
$<$13.8&                      750&                     15.2&      1.01\\
         GQ Lup &                       400&                     17.9&                      950&                     15.3&                      600&                     15.2&    1200\tablenotemark{*}&                  
$<$13.8&     750\tablenotemark{*}&                  $<$14.3&       1.2\\
         IM Lup &      450\tablenotemark{*}&                  $<$17.1&    1100\tablenotemark{*}&                  $<$14.8&                      700\tablenotemark{*}&                  $<$14.8&    1200\tablenotemark{*}&                  
$<$14.1&                      750&                     15.0&      1.11\\
      HD 142527 &      450\tablenotemark{*}&                  $<$17.4&    1100\tablenotemark{*}&                  $<$15.5&                      700\tablenotemark{*}&                  $<$15.1&    1200\tablenotemark{*}&                  
$<$14.3&     750\tablenotemark{*}&                  $<$14.9&      2.97\\
         RU Lup &                       500&                     19.0&    1100\tablenotemark{*}&                  $<$15.2&                      800&                     15.7&    1200\tablenotemark{*}&                  
$<$14.5&                      400&                     16.3&      0.90\\
         RY Lup &      450\tablenotemark{*}&                  $<$17.5&    1100\tablenotemark{*}&                  $<$15.2&                      700\tablenotemark{*}&                  $<$15.2&    1200\tablenotemark{*}&                  
$<$14.2&     750\tablenotemark{*}&                  $<$15.0&      1.21\\
         EX Lup &                       300&                     19.0&                     1600&                     15.3&                      700\tablenotemark{*}&                  $<$14.5&    1200\tablenotemark{*}&                  
$<$13.7&     750\tablenotemark{*}&                  $<$14.3&       2.0\\
         AS 205 &                       300&                     20.6&                      700&                     16.8&                      800&                     15.2&                     1200&                  
   15.3&                      600&                     15.6&       2.1\\
       Haro 1-1 &      450\tablenotemark{*}&                  $<$17.0&    1100\tablenotemark{*}&                  $<$15.0&                      700\tablenotemark{*}&                  $<$14.9&    1200\tablenotemark{*}&                  
$<$13.7&     750\tablenotemark{*}&                  $<$14.6&     0.980\\
       Haro 1-4 &      450\tablenotemark{*}&                  $<$17.2&    1100\tablenotemark{*}&                  $<$15.0&                      700&                     15.0&                     1200&                  
   15.0&     750\tablenotemark{*}&                  $<$14.8&      1.05\\
          VSSG1 &                       350&                     17.8&    1100\tablenotemark{*}&                  $<$14.6&                      600&                     15.0&                      950&                  
   15.6&                      800&                     14.6&       2.1\\
       DoAr 24E &                       450&                     18.3&                     1100&                     15.3&                      800&                     15.8&                     1050&                  
   15.9&                      700&                     15.4&       1.2\\
        DoAr 25 &      450\tablenotemark{*}&                  $<$16.9&    1100\tablenotemark{*}&                  $<$14.9&                      800&                     15.1&                     1200&                  
   14.8&     750\tablenotemark{*}&                  $<$14.6&      1.00\\
          SR 21 &      450\tablenotemark{*}&                  $<$17.4&    1100\tablenotemark{*}&                  $<$15.4&                      700\tablenotemark{*}&                  $<$15.1&    1200\tablenotemark{*}&                  
$<$14.1&     750\tablenotemark{*}&                  $<$14.9&      1.43\\
           SR 9 &      450\tablenotemark{*}&                  $<$17.4&    1100\tablenotemark{*}&                  $<$14.9&                      700\tablenotemark{*}&                  $<$15.1&    1200\tablenotemark{*}&                  
$<$14.1&     750\tablenotemark{*}&                  $<$14.9&      1.07\\
       V853 Oph &                       300&                     18.6&                     1300&                     14.6&                      700&                     14.7&                     1050&                  
   16.0&     750\tablenotemark{*}&                  $<$14.5&       1.2\\
        ROX 42C &      450\tablenotemark{*}&                  $<$17.2&    1100\tablenotemark{*}&                  $<$15.0&                      700\tablenotemark{*}&                  $<$15.0&    1200\tablenotemark{*}&                  
$<$13.9&     750\tablenotemark{*}&                  $<$14.8&      1.05\\
       ROX 43 A &      450\tablenotemark{*}&                  $<$17.4&    1100\tablenotemark{*}&                  $<$14.8&                      700\tablenotemark{*}&                  $<$15.1&    1200\tablenotemark{*}&                  
$<$14.1&     750\tablenotemark{*}&                  $<$14.9&      1.36\\
      Haro 1-16 &                       450&                     18.3&                     1200&                     15.3&                      800&                     15.1&    1200\tablenotemark{*}&                  
$<$14.1&     750\tablenotemark{*}&                  $<$14.7&      0.79\\
      Haro 1-17 &      450\tablenotemark{*}&                  $<$16.8&    1100\tablenotemark{*}&                  $<$14.8&                      700\tablenotemark{*}&                  $<$14.7&    1200\tablenotemark{*}&                  
$<$13.4&     750\tablenotemark{*}&                  $<$14.5&     0.980\\
         RNO 90 &                       450&                     18.3&                     1000&                     15.6&                      750&                     15.5&                     1200&                  
   14.7&                      800&                     15.2&       1.8\\
       Wa Oph 6 &                       400&                     19.3&                      800&                     16.6&                      700&                     15.3&    1200\tablenotemark{*}&                  
$<$14.4&                      500&                     16.0&      0.60\\
      V1121 Oph &                       250&                     20.9&    1100\tablenotemark{*}&                  $<$14.8&                      700\tablenotemark{*}&                  $<$14.9&    1200\tablenotemark{*}&                  
$<$14.0&     750\tablenotemark{*}&                  $<$14.7&       1.7\\
          EC 82 &      450\tablenotemark{*}&                  $<$17.2&    1100\tablenotemark{*}&                  $<$15.0&                      700\tablenotemark{*}&                  $<$14.9&    1200\tablenotemark{*}&                  
$<$14.1&     750\tablenotemark{*}&                  $<$14.7&      1.95\\
\enddata
\tablecomments{Units are cm$^{-2}$ for $N$ and AU for $R$ (where $R=\sqrt{A/\pi}$).}
\tablenotetext{*}{Denotes a fixed temperature for computing upper limits on $N$.}
\end{deluxetable*}

\begin{deluxetable*}{cccccccccccc}
\tablecaption{Model fit results -  HAeBe Systems \label{table:moltable_b}}
\tabletypesize{\scriptsize}
\startdata
Name & T$_\mathrm{H_2O}$ & log N$_\mathrm{H_2O}$ & T$_\mathrm{OH}$ & log  N$_\mathrm{OH}$ &  T$_\mathrm{HCN}$ & log  N$_\mathrm{HCN}$ &  T$_\mathrm{C_2H_2}$ & log N$_\mathrm{C_2H_2}$  &  T$_\mathrm{CO_2}$ &  log N$_\mathrm{CO_2}$ &R \\
       HD 36112 &      450\tablenotemark{*}&                  $<$17.7&    1100\tablenotemark{*}&                  $<$15.6&                      700\tablenotemark{*}&                  $<$15.3&    1200\tablenotemark{*}&                  $<$14.4& 
    750\tablenotemark{*}&                  $<$15.1&      2.13\\
      HD 244604 &      450\tablenotemark{*}&                  $<$17.8&    1100\tablenotemark{*}&                  $<$15.4&                      700\tablenotemark{*}&                  $<$15.5&    1200\tablenotemark{*}&                  $<$14.6& 
    750\tablenotemark{*}&                  $<$15.3&      2.54\\
       HD 36917 &      450\tablenotemark{*}&                  $<$17.9&    1100\tablenotemark{*}&                  $<$15.7&                      700\tablenotemark{*}&                  $<$15.6&    1200\tablenotemark{*}&                  $<$14.7& 
    750\tablenotemark{*}&                  $<$15.4&      2.51\\
       HD 37258 &      450\tablenotemark{*}&                  $<$17.9&    1100\tablenotemark{*}&                  $<$15.4&                      700\tablenotemark{*}&                  $<$15.6&    1200\tablenotemark{*}&                  $<$14.9& 
    750\tablenotemark{*}&                  $<$15.4&      2.90\\
         BF Ori &      450\tablenotemark{*}&                  $<$17.7&    1100\tablenotemark{*}&                  $<$15.4&                      700\tablenotemark{*}&                  $<$15.5&    1200\tablenotemark{*}&                  $<$14.6& 
    750\tablenotemark{*}&                  $<$15.3&      2.30\\
       HD 37357 &      450\tablenotemark{*}&                  $<$17.8&    1100\tablenotemark{*}&                  $<$15.6&                      700\tablenotemark{*}&                  $<$15.5&    1200\tablenotemark{*}&                  $<$14.6& 
    750\tablenotemark{*}&                  $<$15.3&      3.17\\
       HD 37411 &      450\tablenotemark{*}&                  $<$17.7&    1100\tablenotemark{*}&                  $<$15.8&                      700\tablenotemark{*}&                  $<$15.4&    1200\tablenotemark{*}&                  $<$14.4& 
    750\tablenotemark{*}&                  $<$15.2&      2.48\\
         RR Tau &      450\tablenotemark{*}&                  $<$17.7&    1100\tablenotemark{*}&                  $<$15.4&                      700\tablenotemark{*}&                  $<$15.5&    1200\tablenotemark{*}&                  $<$14.5& 
    750\tablenotemark{*}&                  $<$15.2&      13.1\\
       HD 37806 &      450\tablenotemark{*}&                  $<$17.4&    1100\tablenotemark{*}&                  $<$15.1&                      700\tablenotemark{*}&                  $<$15.2&    1200\tablenotemark{*}&                  $<$14.3& 
    750\tablenotemark{*}&                  $<$14.9&      7.19\\
       HD 38087 &      450\tablenotemark{*}&                  $<$18.2&    1100\tablenotemark{*}&                  $<$16.0&                      700\tablenotemark{*}&                  $<$15.9&    1200\tablenotemark{*}&                  $<$14.8& 
    750\tablenotemark{*}&                  $<$15.8&      1.15\\
       HD 38120 &      450\tablenotemark{*}&                  $<$17.6&    1100\tablenotemark{*}&                  $<$15.4&                      700\tablenotemark{*}&                  $<$15.2&    1200\tablenotemark{*}&                  $<$14.4& 
    750\tablenotemark{*}&                  $<$15.0&      7.10\\
       HD 50138 &      450\tablenotemark{*}&                  $<$16.9&    1100\tablenotemark{*}&                  $<$14.6&                      700\tablenotemark{*}&                  $<$14.7&    1200\tablenotemark{*}&                  $<$13.9& 
    750\tablenotemark{*}&                  $<$14.4&      11.5\\
       HD 72106 &      450\tablenotemark{*}&                  $<$17.7&    1100\tablenotemark{*}&                  $<$15.4&                      700\tablenotemark{*}&                  $<$15.4&    1200\tablenotemark{*}&                  $<$14.4& 
    750\tablenotemark{*}&                  $<$15.1&      2.22\\
       HD 95881 &      450\tablenotemark{*}&                  $<$17.3&    1100\tablenotemark{*}&                  $<$14.9&                      700\tablenotemark{*}&                  $<$15.1&    1200\tablenotemark{*}&                  $<$14.2& 
    750\tablenotemark{*}&                  $<$14.9&      1.80\\
       HD 98922 &      450\tablenotemark{*}&                  $<$17.2&    1100\tablenotemark{*}&                  $<$14.8&                      700\tablenotemark{*}&                  $<$15.0&    1200\tablenotemark{*}&                  $<$14.2& 
    750\tablenotemark{*}&                  $<$14.8&      27.5\\
      HD 101412 &      450\tablenotemark{*}&                  $<$17.3&    1100\tablenotemark{*}&                  $<$14.9&                      700\tablenotemark{*}&                  $<$15.1&    1200\tablenotemark{*}&                  $<$14.2& 
                     800&                     16.0&      1.66\\
      HD 144668 &      450\tablenotemark{*}&                  $<$17.3&    1100\tablenotemark{*}&                  $<$15.0&                      700\tablenotemark{*}&                  $<$15.0&    1200\tablenotemark{*}&                  $<$14.2& 
    750\tablenotemark{*}&                  $<$14.8&      4.02\\
      HD 149914 &      450\tablenotemark{*}&                  $<$17.1&    1100\tablenotemark{*}&                  $<$15.0&                      700\tablenotemark{*}&                  $<$14.8&    1200\tablenotemark{*}&                  $<$14.3& 
    750\tablenotemark{*}&                  $<$14.7&      1.00\\
      HD 150193 &      450\tablenotemark{*}&                  $<$17.4&    1100\tablenotemark{*}&                  $<$15.1&                      700\tablenotemark{*}&                  $<$15.1&    1200\tablenotemark{*}&                  $<$14.2& 
    750\tablenotemark{*}&                  $<$14.8&      2.74\\
         VV Ser &      450\tablenotemark{*}&                  $<$17.4&    1100\tablenotemark{*}&                  $<$15.0&                      700\tablenotemark{*}&                  $<$15.1&    1200\tablenotemark{*}&                  $<$14.3& 
    750\tablenotemark{*}&                  $<$14.9&      4.33\\
       LkHa 348 &      450\tablenotemark{*}&                  $<$17.4&    1100\tablenotemark{*}&                  $<$14.9&                      700\tablenotemark{*}&                  $<$15.2&    1200\tablenotemark{*}&                  $<$14.3& 
    750\tablenotemark{*}&                  $<$14.9&      6.12\\
      HD 163296 &      450\tablenotemark{*}&                  $<$17.3&    1100\tablenotemark{*}&                  $<$14.9&                      700\tablenotemark{*}&                  $<$15.0&    1200\tablenotemark{*}&                  $<$14.1& 
    750\tablenotemark{*}&                  $<$14.8&      2.39\\
      HD 179218 &      450\tablenotemark{*}&                  $<$17.3&    1100\tablenotemark{*}&                  $<$15.1&                      700\tablenotemark{*}&                  $<$15.0&    1200\tablenotemark{*}&                  $<$14.1& 
    750\tablenotemark{*}&                  $<$14.8&      5.43\\
      HD 190073 &      450\tablenotemark{*}&                  $<$17.5&    1100\tablenotemark{*}&                  $<$15.1&                      700\tablenotemark{*}&                  $<$15.3&    1200\tablenotemark{*}&                  $<$14.4& 
    750\tablenotemark{*}&                  $<$15.1&      9.19\\
       LkHa 224 &      450\tablenotemark{*}&                  $<$17.2&    1100\tablenotemark{*}&                  $<$14.9&                      700\tablenotemark{*}&                  $<$15.0&    1200\tablenotemark{*}&                  $<$14.1& 
    750\tablenotemark{*}&                  $<$14.8&      15.5
 \enddata
\tablecomments{Units are cm$^{-2}$ for $N$ and AU for $R$ (where $R=\sqrt{A/\pi}$).}
\tablenotetext{*}{Denotes a fixed temperature for computing upper limits on $N$.}
\end{deluxetable*}

A key result of our fits is that the \water\ spectra are fit by a small range of $N$ and $T$, despite the fact that line strengths vary greatly from source to source.   The source-to-source variation in line strength is primarily explained by variations in $A$, with line strength and $A$ being highly correlated.  This is consistent with Figure \ref{fig:outlier_plot}, showing little variation in the \water\ spectra after normalization.   

Our best-fit parameters are similar to those of \citet{Carr08}, although some differences likely arise from their choice of wavelength region and fit minimization.  These choices have non-negligible effects on the model parameters, and there is no `right' choice, since the model is much simpler than the true disk environment.  As an example, if we fit only the region shortward of 14 $\mu$m, we find slightly higher emitting temperatures (by $\sim 100$ K) and smaller emitting areas (by a factor of $\sim$ 2) than the nominal fits,  consistent with the 12--14 $\mu$m lines originating predominately in a disk region closer to the star than the longer wavelength lines.  If we fit the continuum-subtracted spectrum, instead of the individually measured line peaks, 
best-fit models differ from the nominal fit with higher $T$ (by $\sim$100 K) and $N$ (by $\sim$0.5 dex), and smaller $A$ (by $\sim$4$\times$).   Our best-fit $T$ differs more significantly from \citet{Salyk08}, because in that analysis $A$ was fixed at a much smaller value, based on CO lineshapes; $N$ remains similar for both analyses.  

\subsubsection{HCN Results}
Figure \ref{fig:shape_plot_all} shows the peak-normalized mean ({\it after} subtraction of other molecular models) of all HCN-emitting spectra.   The HCN lineshape is fit by a narrow range of temperatures --- 600--800 K --- because at low $T$ the location of the line peak shifts longward, while at high $T$, the model develops a prominent hot-band ``knee'' on the short-wavelength side of the feature.

The underlying H$_2$O features contribute up to $\sim 40$\% of the flux in this region.  Therefore, in the case that the HCN is optically thin, the \water\ model subtraction results in, at most, a 40\% correction in $N$.  However, if $A<<A_\mathrm{H_2O}$ then the emission may be optically thick, and uncertainties in the \water\ flux subtraction could result in larger uncertainties in $N$.  With $A=A_\mathrm{H_2O}$, the best-fit model parameters are shown in Table \ref{table:moltable} and Figure \ref{fig:hist_plot}.   Best-fit column densities range from a few $\times10^{14}$ to a few $\times 10^{15}$ cm$^{-2}$.

There is an unaccounted-for residual in the $\sim$13.8-13.9 $\mu$m region, apparent in Figure \ref{fig:shape_plot_all}, but also visible in many individual spectra.  One possible explanation is that the water has not been modeled correctly, but observed \water\ features do not line up well with the residual.  
The residual does line up with the  $\nu_2 = 5 \rightarrow 4$ and $\nu_2 = 4 \rightarrow 3$ hot-bands of HCN, whose emitting levels are
the lower states of the well-known 130 and 330 $\mu$m submillimeter masers \citep{Gebbie64}, respectively. These high-lying bending states can be
fed either by radiative excitation through combination bands involving the fairly weak C-N ($\nu_1$) stretching mode, or though chemical
reactions that leave the product HCN highly vibrationally excited \citep{Pichamuthu74}. Thus, the presence of this residual suggests that
the bending vibrational levels of HCN may be non-thermally excited, and that caution must be used in the interpretation of the Q-branch
profile using LTE models.

\subsubsection{C$_2$H$_2$ Results}
\label{sec:c2h2}
The mean \acetylene\ emission feature is shown in Figure \ref{fig:shape_plot_all}, after subtraction of both best-fit \water\ models and best-fit HCN models.
With $A=A_\mathrm{H_2O}$, the emission is optically thin, and the suite of emission features is consistent with $T$ in the range 500--1200 K.  Best-fit $N$ range from $10^{14}$ to $10^{15}$ cm$^{-2}$.  The derived $N$ will scale inversely with the assumed $A$ until column densities of $\gtrsim10^{16}\ \mathrm{cm}^{-2}$ when the emission starts to become optically thick.

In addition to the residual at $\sim$ 13.85 $\mu$m, three residual peaks shortward of 13.6 $\mu$m are prominent in many sources, and their origin is unknown.  At least one peak lines up with an \water\ feature, and may be accounted for with a better \water\ model.  The 13.55 $\mu$m feature is at the location of a high-temperature OH emission line, and could be evidence for highly-excited OH, like that seen in TW Hya \citep{Najita10}.  However, adjacent transitions in the OH rotational ladder are not apparent in the spectrum, and so this may not be the correct explanation.  

\subsubsection{CO$_2$ Results}
The mean CO$_2$ feature is shown in Figure \ref{fig:shape_plot_all}.  The emission is mostly optically thin with $A=A_\mathrm{H_2O}$, though slightly optically thick at the center of the Q-branch peak.  The complete suite of feature shapes is consistent with $T$ from 500--1200 K, but best-fit $T$ cluster near 700 K.  Best-fit $N$ are $\sim10^{14}-10^{16}$ cm$^{-2}$ and are inversely proportional to $A$ until $N\gtrsim10^{17}$ cm$^{-2}$.  

No significant residuals are noted; those seen in Figure \ref{fig:shape_plot_all} are most likely \water\ model subtraction residuals or data-reduction artifacts.

\subsubsection{OH Results}
OH was fit with emission doublets from 17 to 30.5 $\mu$m.  Line strengths decrease towards shorter wavelengths.  Derived excitation temperatures are strongly dependent on the strength of the relatively weak short wavelength lines, which are in turn dependent on the continuum subtraction, making $T$ somewhat more uncertain for OH than for other molecules.   In addition, with $A=A_\mathrm{H_2O}$, the low energy lines are beginning to become optically thick, and so the shape of the emission spectrum is influenced by both $N$ and $T$.

\begin{figure*}
\epsscale{1}
\plotone{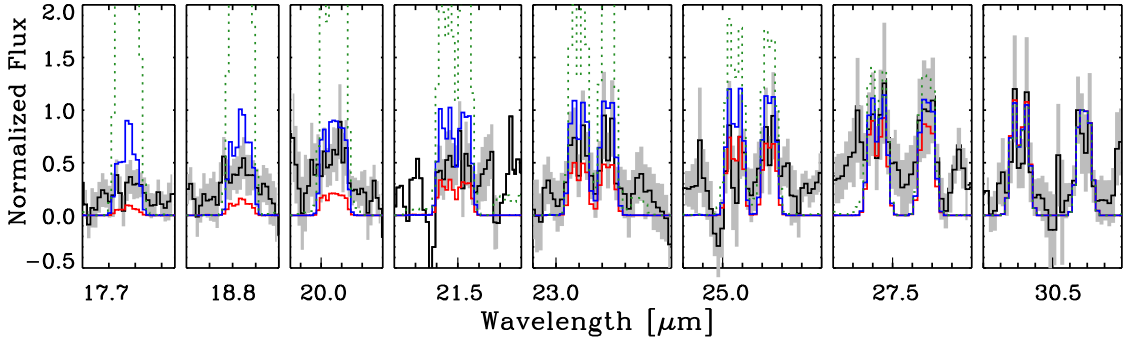}
\caption{Mean (black solid line) plus or minus the standard deviation (gray) of normalized OH emission sources, after subtraction of other molecular emission models. Plotted in red, blue, and dotted green, respectively, are optically thin 700, 1600, and 4000 K models, normalized to the 30.5 $\mu$m peaks.  
 \label{fig:shape_plot_oh}}
\end{figure*}

Figure \ref{fig:shape_plot_oh} shows the rather large range of $T$ (700--1600 K) that may be appropriate for fitting OH, with the mean profile in this case composed of spectra normalized to the 30.7 $\mu$m line peak.  A much higher $T$ is inconsistent with the data, suggesting a difference between the emission observed here primarily in the Long-High module, and the OH emission observed in the Short-High module in the transitional disk TW Hya \citep{Najita10}, though the latter preferentially probe lines with higher excitation energy.  Potential evidence for higher excitation lines in our data is the residual line at 13.55 $\mu$m (discussed in \S \ref{sec:c2h2}) which is coincident with a high-excitation OH transition.  High excitation temperatures may be indicative of non-thermal excitation, such as in the case of the ``prompt'' emission observed in comets \citep[e.g.][]{Bonev06}, in which rotationally-excited OH is produced via the photolysis of \water.  For the most part,  the moderate excitation temperatures we observe are instead reasonable thermal temperatures for few-AU disk atmospheres around cTTs disks.

Best-fit temperatures are $>$800 K and typically lie between 900 and 1100 K.  Best-fit $N$ are near $10^{15}$ cm$^{-2}$.  The strongest lines are already becoming optically thick at these column densities, so $N$ could differ significantly if $A\neq A_\mathrm{H_2O}$.  For example, in a test with the emitting area reduced by a factor of 10, we find the derived $N$ to be increased by about a factor of 25.  Similarly, \citet{Salyk08} found column densities of $2\times10^{17}$ cm$^{-2}$ when fitting 3 $\mu$m OH emission features for DR Tau and AS 205, which are a factor of $\sim$50 higher than those found in this analysis; however, emitting areas were fixed to values 27 and 30 times smaller, respectively.  These $N$ are significantly smaller than those found by \citet{Bethell09} to be consistent with water self-shielding, but we note that higher $N$ of the right order for this process are not ruled out with smaller $A$.

\subsubsection{CO Results}
Best-fit model parameters can be seen in Table \ref{table:co_fit_parameters} and Figure \ref{fig:hist_plot}.  Temperatures are typically $\gtrsim900$ K, with a cluster at very high $T$ ($\gtrsim 1500$ K), though temperatures this high may be unrealistic.  Despite the large range of $T$, $N$ ranges over only about 1 dex, from $\sim$10$^{18}-10^{19}\ \mathrm{cm}^{-2}$.  

\begin{deluxetable}{cccccccccccc}
\tablecaption{CO Model Fit Results \label{table:co_fit_parameters}}
\startdata
Name & T$_\mathrm{CO}$ & log N$_\mathrm{CO}$ & R$_\mathrm{CO}$ \\
  LkHa 326  &           1550  &       19.3  &       0.08  \\
  LkHa 327  &            1625  &       19.5  &       0.10  \\
  LkHa 330  &            850  &       17.4  &       0.30  \\
    LkCa 8  &          1475  &       14.6  &       2.41  \\
  V710 Tau  &         1675  &       19.3  &       0.05  \\
    AA Tau  &            950  &       18.6  &       0.13  \\
    DR Tau  &           1250  &       18.9  &       0.17  \\
HD 135344 B  &            900  &       17.9  &       0.10  \\
    GQ Lup  &           1675  &       18.6  &       0.08  \\
    AS 205  &         1000  &       18.7  &       0.35  \\
  Haro 1-4  &        1675  &       18.5  &       0.04  \\
     VSSG1  &         1600  &       18.6  &       0.09  \\
  DoAr 24E  &        1275  &       18.3  &       0.21  \\
      SR 9  &          1575  &       16.9  &       0.16  \\
  V853 Oph  &           975  &       18.0  &       0.09  \\
 Haro 1-16  &           1250  &       18.2  &       0.08  \\
    RNO 90  &           1625  &       18.2  &       0.22  \\
  Wa Oph 6  &          1675  &       18.5  &       0.10  \\
 V1121 Oph  &         1125  &       18.5  &       0.12  \\
     EC 82  &           1000  &       17.9  &       0.39  \\
  HD 36112  &           775  &       17.9  &       0.51  \\
  HD 37357  &            575  &       19.5  &       0.61  \\
  HD 37806  &            625  &       19.8  &       1.39  \\
  HD 38120  &         800  &       18.3  &       0.27  \\
  HD 50138  &            725  &       19.1  &       0.97  \\
 HD 150193  &             725  &       18.5  &       0.47  \\
    VV Ser  &        1175  &       18.1  &       0.61  \\
 HD 163296  &            850  &       18.5  &       0.32  \\
 HD 190073  &           525  &       20.4  &       2.64  \\
\enddata
\tablecomments{Units are cm$^{-2}$ for $N$ and AU for $R$ (where $R=\sqrt{A/\pi}$).}
\end{deluxetable}

\subsection{Molecular Ratios relative to \water}
\label{section:molratio_water}
\begin{figure}
\epsscale{1}
\plotone{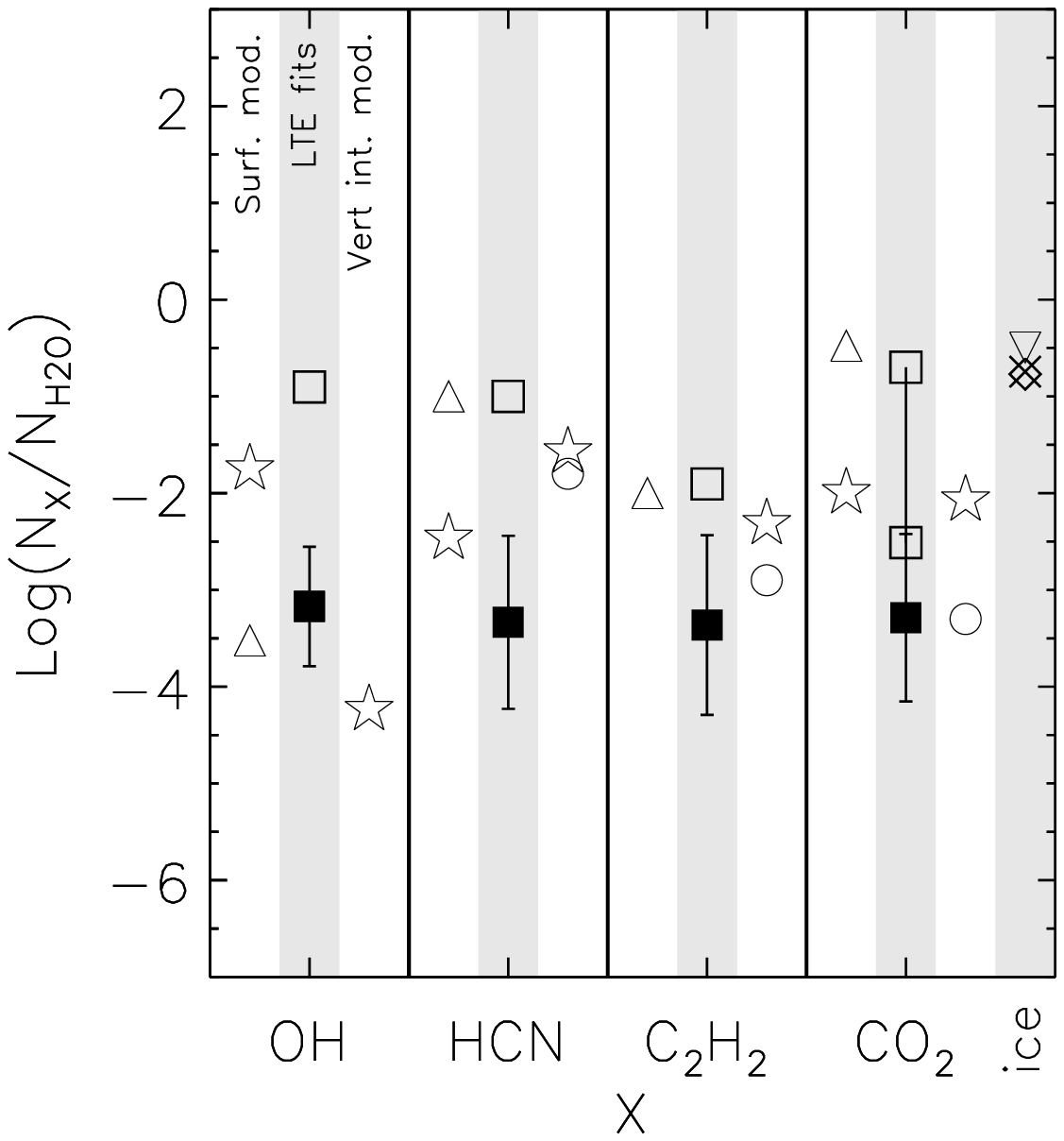}
\caption{Molecular ratios, relative to H$_2$O (filled squares), with error bars showing the 1$\sigma$ scatter. Also shown are previous observations in disks \citep[open squares:][]{Carr08}, quiescent cloud ices \citep[cross:][]{Whittet07}, high-mass YSO ices \citep[open diamond:][]{Gerakines99} and low-mass YSO ices \citep[open inverted triangle:][]{Pontoppidan08b}.  To the left are models of the disk surface \citep[open stars:][]{Willacy09} and photon-dominated regions \citep[open triangles:][]{Agundez08}; to the right are vertically integrated model values at 1 AU (open stars: Willacy \& Woods 2009; circles: Markwick et al. 2002).
 \label{fig:molratio_plot}}
\end{figure}

Figure \ref{fig:molratio_plot} shows molecular ratios with respect to \water. $N_X/N_\mathrm{H_2O}$$\sim10^{-3}$ for all molecules.  Also plotted are ratios derived from previous disk observations \citep{Carr08}, ice observations in 
quiescent clouds \citep{Whittet07}, high-mass YSOs \citep{Gerakines99} and low-mass YSOs \citep{Pontoppidan08b}, as well as several disk models \citep{Markwick02, Agundez08, Willacy09}.  Additionally, \citet{Glassgold09} (not plotted) predict OH/\water\ ratios in the warm molecular layer of the disk similar to the ratios found here --- $10^{-4}-10^{-2}$ for a variety of chemical models, sampled at radii from 0.25--3 AU.  

The large differences between our work and that of \citet{Carr08} arise from differences in derived column densities for both \water\ and any particular molecule, which are due to differences in choice of fitting method and assumptions, as discussed above.  With Spitzer-IRS, the Q-branch transitions are completely blended together, but at higher spectral resolution individual Q, P, or R-branch lines could be resolved, which would make possible independent determinations of emitting area.  This highlights the need for higher-spectral-resolution and, eventually spatial resolution, observations of molecular emission.  The brightest sources are being pursued at high resolution with current instrumentation \citep[e.g.][]{Pontoppidan10b}, but a more complete census of water-emitting disks may require the next generation of ground- and space-based mid-IR spectrographs.

CO$_2$/\water\ ratios in ices have been observed to be $\sim 0.1-1$ in a variety of environments \citep{Gerakines99, Whittet07,Pontoppidan08b}, higher than our observed values and at the upper range of those derived by \citet{Carr08} (see the far right of Figure \ref{fig:molratio_plot}).    This may reflect chemical evolution towards lower values in the inner disk region, as seen, for example, in the models of \citet{Markwick02}.

Figure \ref{fig:molratio_plot} also shows molecular ratios derived from thermo-chemical disk models. Disk-integrated column density ratios at 1 AU from \citet{Markwick02} (circles) and \citep{Willacy09} (stars) are plotted to the right side of the observations, while column density ratios from the disk surface \citep[T$>$150 K; stars;][]{Willacy09} and from photo-dominated regions (PDR) modeled by \citet{Agundez08} (triangles) are plotted to the left.  Although we show ratios based on integrated vertical columns, these are likely not relevant for the disk surface being probed by the IRS molecular emission.  

It is immediately apparent that comparisons of data and models can be quite complex.  For example, orders of magnitude differences in molecular ratios are computed when considering different regions of the disk, both radially and vertically, for many molecules.  And, the emitted lines themselves originate from a range of disk radii and heights, which are not even the same from transition to transition.  For example, \citet{Woitke09} find that pure rotational \water\ lines could in principle originate in both the inner and outer ($>$10 AU) disk, and even inside the disk inner rim.  Nevertheless, the scatter observed in column density ratios is smaller than some of the model differences, holding promise for observations to eventually distinguish between models.   One thing we can already say is that the observed abundances of \acetylene\ require active carbon chemistry, such as that described in \citet{Agundez08}, as models without such networks produce nowhere near enough \acetylene\ in the inner disk atmosphere; for example, \citet{Willacy09} find \acetylene/\water\ ratios of $\sim10^{-10}$ (not plotted).  Alternatively, the high
abundance of \acetylene\ could be produced by the burning of polycyclic aromatic hydrocarbons (PAHs; though the production of PAHs may require \acetylene\ as a precursor; see discussion in \S \ref{sec:pah}).

A more promising approach to comparing models and observations may be to couple thermo-chemical models with line-generating radiative transfer code, and directly compare output model spectra with observations.  This approach has been pursued, for example, by \citet{Kamp10} for atomic fine-structure lines and low-J rotational CO lines, as well as \citet{Meijerink08} and \citet{Woitke09} for far-IR rotational lines of \water.  An alternative approach is to map abundance structures using observations and radiative transfer, and compare these to the abundances predicted by various thermo-chemical models.  This approach has already yielded interesting results about water abundances \citep{Meijerink09}, and we expect that it will lead to significant progress in the comparison of chemical models and observations in the near future.

\subsection{Molecular ratios relative to CO}
\begin{figure}
\epsscale{1}
\plotone{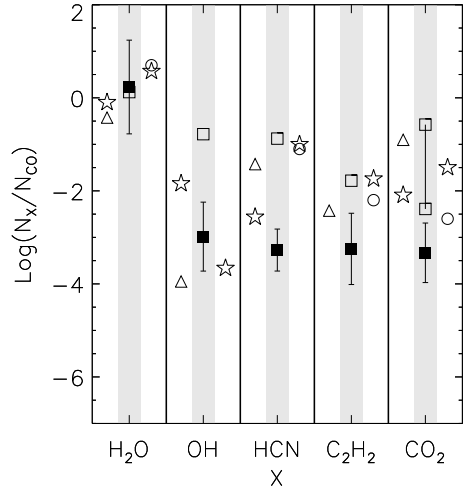}
\caption{Molecular ratios, relative to CO.  References and symbols are the same as in Figure \ref{fig:molratio_plot}.
 \label{fig:co_ratio_plot}}
\end{figure}

In Figure \ref{fig:co_ratio_plot}, we show molecular ratios with respect to CO.   We find \water/CO ratios slightly above 1 --- nearly the same as that found by \citet{Carr08} and lying between vertically-integrated and disk surface model ratios.  \citet{Glassgold09} also find \water/CO ratios near 1 for several types of models.  Again, other molecular ratios are different from those found in \citet{Carr08} because of our choice of modeling assumptions.  The scatter in \water/CO ratio is only $\sim$0.5-1 dex and so these ratios may prove helpful in distinguishing between chemical models.  Note, however, that we have found significantly different emitting areas for CO and \water\, and so the NIRSPEC and IRS emission lines are almost certainly probing different emitting regions.   

\section{Summary of Results}
We have presented and analyzed Spitzer-IRS spectra of \water, OH, HCN, \acetylene\ and CO$_2$ and Keck-NIRSPEC spectra of CO from a large sample of cTT and HAeBe disks.
In addition to the detection dependence on spectral type and SED class (transitional versus classical) noted in Paper I, we also find that detection efficiency depends on the disk color, $n_{13-30}$, H$\alpha$ equivalent width and, tentatively, on $\dot{M}$.  Radiative transfer disk models suggest a number of ways to enhance line emission, including increasing the molecular abundance, lowering the dust/gas ratio in the disk atmosphere, and increasing gas heating \citep{Meijerink09}.  Since $n_{13-30}$ is a measure of the settling of small grains in the disk atmosphere, these data suggest that it may be necessary to lower the dust/gas ratio in the atmosphere in order to produce these lines.  In addition, the correlation with H$\alpha$ suggests that accretion is also important, perhaps because it provides additional heating for the excitation of lines.

We find that detections, line fluxes and line-to-continuum ratios for nearly all molecules are correlated, suggesting a common origin or common excitation conditions for this forest of lines.  Line fluxes (when detected) are also correlated with continuum flux and many other covarying system parameters (including L$_\star$, M$_\star$ and $\dot{M}$), and once this variation is accounted for, no other statistically significant correlations are noted between line fluxes and system parameters.  The observed correlation could be the result of any of these physical parameters, but a good candidate is the stellar luminosity, which provides heating for the line emission, and will also determine the size of the line-emitting region.

Although line fluxes are correlated, line flux ratios (for example, HCN/\water) do vary from source to source.  No obvious trends in line flux ratios are seen, but the variations suggest real chemical differences whose origin may be revealed with future study.

We also compare the mid-IR emission lines with near-IR CO vibrational emission observed with Keck-NIRSPEC. Interestingly, CO lines are often seen even when mid-IR emission is not, including in many HAeBe and transitional disks.  However, rather than find evidence for different \water/CO ratios, we instead find 
that both transitional disks and HAeBe disks have systematically lower CO line/continuum ratios, consistent with the reduced IRS line/continuum ratios reported in Paper I.  This suggests possible systematic differences in line excitation between the two types of sources.  

While \water\ line strengths do differ significantly from source to source, the emission spectra are similar if normalized by line strength.  In the context of an LTE model, the line strength is primarily determined by the emitting area, and so this accounts for most of the variation.  Temperature and column density are then determined by the spectral slope.  Best-fit temperatures lie near 450 K and N$\sim10^{18}\,\mathrm{cm}^{-2}$ for all sources, though results depend somewhat on the choice of fitting method and wavelength region.  The high temperatures suggest an origin in the planet-forming (few AU) region of the disk, and $N$ is consistent with chemical models of \water\ formation \citep[e.g.][]{Glassgold09} as well as with models that include UV destruction and \water\ self-shielding \citep{Bethell09}.  

Other molecules are fit with a range of excitation temperatures, but all are representative of the planet-forming region of the disk, some 500 K or higher.  Because of model degeneracies, coupled with systematic uncertainties, we chose to fix emitting areas, but emitting areas and column densities can in the future be scaled to match results from additional observations.  With this technique, we derive molecular ratios with respect to water and CO of $\sim10^{-3}$ for HCN, \acetylene, CO$_2$ and OH.  We find CO/\water\ ratios near 1, similar to values from previous observations \citep{Carr08} and several disk models \citep{Markwick02,Agundez08,Glassgold09,Willacy09}. It is not entirely straightforward to relate the column densities derived here to those derived from thermo-chemical disk models.  However, the small scatter in observed molecular column density ratios, compared to large differences between models, suggests that comparisons between these data and output of line-generating thermo-chemical codes holds great promise for pinpointing the most important physical and chemical processes in protoplanetary disks.  

\section{Discussion}
\subsection{Completeness of the observed chemistry}
Given the detections of a wide range of abundant molecules containing three primary elemental
constituents of disks, oxygen, carbon and nitrogen, 
it is possible to compare the observed abundances of these species to the total expected elemental
abundance, assuming solar values. 
Obviously, not all molecular species have been observed, but what fraction of them are we likely to have seen? Using the
average column density values
from Figure \ref{fig:hist_plot}, we have accounted for total column densities of $\rm O\sim 4\times10^{18}\, \mathrm{cm}^{-2}$, $\mathrm{C}\sim 1\times10^{18}\,
\mathrm{cm}^{-2}$, and $N\sim 2\times10^{15}\, \mathrm{cm}^{-2}$, 
corresponding to ratios of $\mathrm{O/C}\sim 4$ and $\mathrm{N/O}\sim 5\times10^{-4}$. Comparing this to the solar ratios of $\mathrm{
O/C}_{solar}=1.6$ and $\rm N/O=0.15$ \citep{Frisch03}, 
it is apparent that carbon is relatively underrepresented, while we only detect a very small fraction of the available
nitrogen. The uncertainties in
the derived abundances, along with the possibility that the different species do not probe the same radii, probably
preclude any strong conclusions to be drawn from the derived O/C ratio, except to note that it is consistent with the
observed oxygen-dominated chemistry rich in H$_2$O. The missing nitrogen, however, seems to indicate that a significant
chemical reservoir still remains to be discovered. Possible carriers of the remaining nitrogen are NH$_3$ and N$_2$. 

NH$_3$ emits strongly near 10$\mu$m, so Spitzer-IRS can be used to place limits on its column density.  With a conservatively
low temperature (400 K) and assuming the same emitting area as that for the \water\ fits, we find upper limits near 10$^{16}\ \mathrm{cm}^{-2}$, or only 
about 10$\times$ the observed column density of HCN.  Therefore, NH$_3$ can not account for the orders of magnitude 
of missing N.  A smaller N deficiency has been noted in cometary observations \citep{Jessberger91, Wyckoff91a}, with one
explanation being that N$_2$ is the primary reservoir for N in the solar nebula, and is not easily retained by comets \citep{Iro03}.  Thus, N$_2$ may be a likely reservoir for the disks observed in this study as well.

\subsection{The column density of the molecular layer}
The elemental abundances discussed above place strict lower limits on the total column density of the warm molecular layer
at $\sim$1\,AU. The strongest constraint is
given by oxygen, even with a significantly unconstrained atomic O component. Using the ISM oxygen abundance, the thickness
of the warm molecular layer must be at least $N(H)\sim 7\times10^{21}\,\rm cm^{-2}$. This is a
very good match to existing models \citep[e.g.][]{Glassgold09}.

\subsection{Implications for the carbon deficit in the inner solar system}
The terrestrial planets in the solar system have a well-known deficit of carbon, by as much as three orders of magnitude
relative to the elemental abundances in the sun. 
Observations of polluted (by accreting debris) white dwarf atmospheres also indicate similar carbon deficits in other
planetary systems \citep{Jura08}. 
The high abundances of carbon observed in the inner disk surfaces apparently do not allow for a similar carbon deficit;
while absolute C and O abundances (relative to H) may remain uncertain, a carbon deficit of a factor $10^3$ would
imply that the column densities of warm molecular gas are very large, at least $N$(H)$\sim N$(CO)$/(10^{-3} x($C$) f) =
3/f\times10^{24}\,\mathrm{cm}^{-2}$, where $f$ is the fraction of C accounted for in the observed lines, and that a
comparatively small fraction of the elemental oxygen is accounted for in the observations. Both of these requirements
seem implausible in light of recent thermo-chemical models. More likely, the observed high abundance of carbon
in the gas phase indicates that only a small fraction of it will eventually be incorporated into terrestrial planets (however, see the discussion of the formation of refractory organics in \S \ref{sec:pah}). 

Recently, \cite{Lee10}
proposed that primordial carbon grains are destroyed in the disk surface through chemisputtering by free oxygen atoms;
essentially they burn, producing CO gas. If vertical mixing is efficient, these authors suggest that this process can
reduce inner disk carbon to the observed solar system depletion levels as the CO gas dissipates. We do observe the bulk of the
gas-phase carbon in the form of CO, a significant fraction of which could be formed by destruction of carbon grains. That
there is no obvious observed carbon deficit in the gas phase is fully consistent with the picture proposed by \cite{Lee10}. If
all of the relatively volatile gas-phase CO is removed, and assuming there is no carbon left in primordial grains, the
carbon remaining in CO$_2$, C$_2$H$_2$ and HCN is consistent with a carbon deficit of $10^{-2}-10^{-3}$. The main uncertainty is likely the
abundance of CH$_4$, which does not have emission bands in the range of the Spitzer IRS high resolution modules. Non-detection
of free carbon may indicate that it is not abundant in the disk surface \citep{Mathews10}.

\subsection{Abundant acetylene and aromatic hydrocarbons}
\label{sec:pah}
Many protoplanetary disks are known to harbor PAHs, at least in their surfaces
\citep{Habart04, Geers07a}, and some of them have emission spectra entirely dominated by the characteristic mid-infrared
PAH emission bands \citep{Geers07b}. While the detection fraction is high in HAeBe stars, the relative
lack of exciting UV photons in disks around young solar analogs and lower mass stars makes positive detections difficult
even if PAHs are actually present in abundance \citep{Geers07a}. The observed PAHs are likely analogs to the solar nebula
precursors of the aromatic component of carbonaceous chondritic meteorites \citep{Pering71}, which, due to their
refractory properties, may represent an important route for delivery of organics to the surfaces of terrestrial planets.
The pathway to PAH presence in disks, however, is still debated.

The formation of aromatics by hydrocarbon combustion chemistry in the atmospheres of asymptotic giant branch stars was
proposed by \cite{Frenklach89} and quantified using a
detailed kinetic model. The model was applied to the solar nebula by \cite{Morgan91} to explain the abundance of aromatics
in primitive chondritic meteorites. Their finding is that the formation of PAHs is predicated on (1) the presence of dense
gas at temperatures of 900-1100\,K and (2) a high concentration of acetylene, C$_2$H$_2$. Their conclusions were, however,
speculative since it was not known whether the necessary precursor, \acetylene, was indeed present in the early solar
nebula. 

Our finding that C$_2$H$_2$ is a common constituent of protoplanetary disk surfaces at temperatures of 600--1200\,K
therefore has direct implications for the synthesis of PAHs within typical protoplanetary disks, including the solar
nebula. \cite{Morgan91} find that the PAH yield (fraction of initial carbon atoms incorporated into PAH molecules) is a
sensitive function of pressure as well as the initial C$_2$H$_2$ concentration. The most favorable pressures correspond to
gas densities of $\sim$$\rm 10^{14}-10^{15}\,cm^{-3}$ and high abundances of C$_2$H$_2$/H$_2$$\sim$$10^{-6}-10^{-5}$,
leading to PAH yields (the fraction of C atoms in the initial gas incorporated into PAHs) of order unity. The observed \acetylene\ abundances are 1--2 orders of magnitude lower. However, actual
PAH yields may only be of order a few \%, and a significant part of the C$_2$H$_2$ may already have been converted to PAHs
at the typical age of our disk sample (1--3\,Myr).  Similarly, at sufficiently high abundances HCN can polymerize into purine
and pyrimidine compounds (such as those present in nucleic acids), and thus the observation of abundant C$_2$H$_2$ and HCN at 1000\,K
supports the possibility of a complex organic/``soot ring'' chemistry in protoplanetary disks, which may be revealed by high resolution imaging,
including mid-IR interferometry. 

Alternatively, it has been shown that starting with abundant PAH's, chemical reactions with H, OH and O tend to destroy PAHs in the
terrestrial planet-forming regions of disks \citep{Kress10}.  One by-product of this process is \acetylene, with a maximum 
\acetylene/CO ratio of $\sim10^{-2}$.  In either case, the region of the disk surface traced by the mid-infrared molecular line forest may
indeed be an important site for understanding the synthesis and/or destruction of the carbonaceous, aromatic component of chondrites.

\subsection{Conclusion}
This work presents an overview of the rich near- and mid-IR molecular emission observed from protoplanetary disks.
What physical mechanisms account for the presence or absence of molecular lines, and for differences in molecular line ratios from source to source?  What physical process are most crucial for determining chemical abundances in the inner disk, and is there inheritance of information from the cloud it formed out of? We have begun to answer these questions, 
showing here and in Paper I the importance of photochemistry, disk heating, grain settling and various chemical processes in the production of molecular line emission.
However, it is also clear that there are variations within the sample that have yet to be explained.   In the near future, these questions will continue to be 
explored, with radiative transfer codes and thermo-chemical disk models.
We also expect that the data presented here will provide a rich database for use with observations obtained both in the near future and beyond, with facilities such as the Herschel Space Observatory, ALMA, SOFIA, JWST, and the emerging class of `extremely large' telescopes.
\newline
\newline
\noindent The authors would like to thank Eric Feigelson for useful discussions about the connection between C$_2$H$_2$ and PAHs.  This work is based on observations made with the Spitzer Space Telescope, which is operated by the Jet Propulsion Laboratory, California Institute of Technology under a contract with NASA. Support for this work was provided by NASA. Some of the data presented herein were obtained at the W.M. Keck Observatory, which is operated as a scientific partnership among the California Institute of Technology, the University of California and the National Aeronautics and Space Administration. The Observatory was made possible by the generous financial support of the W.M. Keck Foundation.  Support for KMP was provided by NASA through Hubble Fellowship grant no. 01201.01 awarded by the Space Telescope Science Institute, which is operated by the Association of Universities for Re- search in Astronomy, Inc., for NASA, under contract NAS 5-26555. Research support for JSC was also provided by 6.1 base funding at the Naval Research Laboratory.

\end{document}